\documentclass [11pt,a4paper]{article}
\usepackage{jheppub}

\usepackage{hyperref}
\usepackage[english]{babel}
\usepackage[latin1,utf8]{inputenc}
\usepackage[OT2,T1]{fontenc}
\usepackage{graphicx}
\usepackage{color}
\usepackage{esint}
\usepackage{bm,bbm}
\usepackage{pifont} 
\usepackage{changepage}

\usepackage{amsmath,amsfonts,amsthm,amssymb}
\usepackage{tikz,stackrel}
\newcommand*\circled[1]{\tikz[baseline=(char.base)]{
            \node[shape=circle,draw,inner sep=1pt] (char) {#1};}}
\newcommand*\squared[1]{\tikz[baseline=(char.base)]{
            \node[shape=rectangle,draw,inner sep=2pt] (char) {#1};}}
\newcommand\cyr{%
\renewcommand\rmdefault{wncyr}%
\renewcommand\sfdefault{wncyss}%
\renewcommand\encodingdefault{OT2}%
\normalfont
\selectfont}
\DeclareTextFontCommand{\textcyr}{\cyr}

\newcommand{\be}{\begin{equation}}
\newcommand{\ee}{\end{equation}}
\newcommand{\ba}{\begin{eqnarray}}
\newcommand{\ea}{\end{eqnarray}}

\newcommand{\no}{\nonumber\\}

\def\bs{\begin{subequations}}
\def\es{\end{subequations}}
\def\a{\alpha}
\def\b{\beta}
\def\de{\delta}
\def\g{\gamma}
\def\G{\Gamma}
\def\la{\lambda}
\def\k{\kappa}

\def\G{\Gamma}
\def\t{\tau}  
\def\s{\sigma}

\def\vp{\varphi}
\def\N{\nabla}

\def\cF{\mathcal{F}}

\def\cK{\mathcal{K}}
\def\cL{\mathcal{L}}

\def\cR{\mathcal{R}}
\def\cS{\mathcal{S}}

\def\p{\partial}

\def\B{\Box}
\newcommand{\Eq}[1]{(\ref{#1})}
\def\com{\color{magenta}}
\def\cob{\color{blue}}

\newcommand{\oarX}[1]{\href{http://arxiv.org/abs/#1}{{\ttfamily\com arXiv:#1}}}
\newcommand{\arX}[1]{\href{http://arxiv.org/abs/#1}{{\ttfamily\com arXiv:#1}}}
\newcommand{\doin}[6]{\href{http://dx.doi.org/#1}{{\cob {\it #2} {\bf #3 #4} (#6) #5}}}
\newcommand{\doinn}[5]{\href{http://dx.doi.org/#1}{{\cob {\it #2} {\bf #3} (#5) #4}}}
\newcommand{\doij}[5]{\href{http://dx.doi.org/#1}{{\cob {\it #2} {\bf #3} (#5) #4}}}

\newcommand{\ndoinn}[5]{\href{#1}{{\cob {\it #2} {\bf #3} (#5) #4}}}

\newcommand{\tia}[1]{\textit{#1},}
\newcommand{\boxd}[1]{\boxed{\phantom{\Biggl(}#1\phantom{\Biggl)}}}

\def\rme{e}
\def\rmd{d}
\def\rmi{i}
\def\tfrac{\textstyle{\frac12}}

\begin{document}

\title{Initial conditions and degrees of freedom of non-local gravity}

\author{Gianluca Calcagni,}
\emailAdd{g.calcagni@csic.es}
\affiliation{Instituto de Estructura de la Materia, CSIC, Serrano 121, 28006 Madrid, Spain}

\author{Leonardo Modesto,}
\emailAdd{lmodesto@sustc.edu.cn}
\affiliation{Department of Physics, Southern University of Science and Technology, Shenzhen 518055, China} 

\author{Giuseppe Nardelli}
\emailAdd{giuseppe.nardelli@unicatt.it}
\affiliation{Dipartimento di Matematica e Fisica, Universit\`a Cattolica del Sacro Cuore,\\ via Musei 41, 25121 Brescia, Italy}
\affiliation{TIFPA -- INFN c/o Dipartimento di Fisica, Universit\`a di Trento,\\ 38123 Povo (Trento), Italy}

\abstract{We prove the equivalence between non-local gravity with an arbitrary form factor and a non-local gravitational system with an extra rank-2 symmetric tensor. Thanks to this reformulation, we use the diffusion-equation method to transform the dynamics of renormalizable non-local gravity with exponential operators into a higher-dimensional system local in spacetime coordinates. This method, first illustrated with a scalar field theory and then applied to gravity, allows one to solve the Cauchy problem and  count the number of initial conditions and of non-perturbative degrees of freedom, which is finite. In particular, the non-local scalar and gravitational theories with exponential operators are characterized by, respectively, two and four initial conditions in any dimension and, respectively, by one and eight degrees of freedom in four dimensions. The fully covariant equations of motion are written in a form convenient to find analytic non-perturbative solutions.}

\date{March 1, 2018}

\preprint{\doij{10.1007/JHEP05(2018)087}{JHEP}{05}{087}{2018}; \doij{10.1007/JHEP05(2019)095}{JHEP}{05}{095}{2019} [\arX{1803.00561}]}

\maketitle

\tableofcontents


\section{Introduction}\label{intro}

There is cumulative evidence that theories with exponential non-local operators of the form
\be\label{serep0}
\rme^{-(\B/M^2)^n}
\ee
have interesting renormalization properties. After early studies of quantum scalar field theories \cite{AE1,AE2,Efi77,Efi01} and gauge and gravitational theories \cite{Kra87,Mof1,HaMo,EMKW,Cor1,Cor2,Cor3}, in recent years there has been a surge of interest in non-local classical and quantum gravity \cite{Kuz89,Tom97,Bar1,Bar2,BMS,Kho06,cuta8,BKM1,Mof3,Mod1,BGKM,Mod2,Mod3,Mod4,BMTs,BCKM,CaMo2,MoRa1,TBM,MoRa2,MoRa3,Edh18,BKLM}. A non-local theory of gravity aims to fulfill a synthesis of minimal requirements: (i) spacetime is a continuum where Lorentz invariance is preserved at all scales; (ii) classical local (super-)gravity should be a good approximation at low energy; (iii) the quantum theory must be perturbatively super-renormalizable or finite; (iv) the quantum theory must be unitary and ghost free, without extra pathological degrees of freedom in addition to those present in the classical theory; (v) typical classical solutions must be singularity-free.

The typical structure of the gravitational action in $D$ topological dimensions is
\be\nonumber
S_g = \frac{1}{2\kappa^2} \int \rmd^D x \sqrt{-g}\,\left[R-2\Lambda + R_{\mu \nu} \, \cF_2(\B) \, R^{\mu \nu} + R  \cF_0(\B) R\right],
\ee
where $\k^2=8\pi G$ is the gravitational constant and $\cF_{0,2}$ are \emph{form factors} dependent on the dimensionless ratio $r_*\B:=\B/M^2$, where $M=1/\sqrt{r_*}$ is the characteristic energy scale of the system, $\B=\N_\mu \N^\mu$ is the Laplace--Beltrami or d'Alembertian operator and $\N_\nu V_\mu := \p_\nu V_\mu-\G^\s_{\mu\nu}V_\s$ is the covariant derivative of a vector $V_\mu$. Our conventions for the curvature invariants are
\ba
&& \Gamma^\rho_{\mu\nu}:= \frac12 g^{\rho\s}\left(\p_{\mu} g_{\nu\s}+\p_{\nu} g_{\mu\s}-\p_\s g_{\mu\nu}\right)\,,\label{leci}\\
&& R^\rho_{~\mu\sigma\nu}:= \p_\sigma \Gamma^\rho_{\mu\nu}-\p_\nu \Gamma^\rho_{\mu\sigma}+\Gamma^\tau_{\mu\nu}\Gamma^\rho_{\sigma\tau}-\Gamma^\tau_{\mu\sigma}\Gamma^\rho_{\nu\tau}\,,\label{rite}\\
&& R_{\mu\nu}:= R^\rho_{~\mu\rho\nu}\,,\qquad R:= R_{\mu\nu}g^{\mu\nu}\,.
\ea
The particular choice of form factors
\be
\cF_2(\B) = \frac{\rme^{-r_*\B} -1}{\B} \,,\qquad \cF_0(\B) = -\frac{\rme^{-r_*\B}-1}{2\B} \,,\nonumber
\ee
leads to the action \cite{Mod1,Mod2,Mod3,Mod4,CaMo2}
\be
S_g = \frac{1}{2\kappa^2}\int \rmd^D x \sqrt{-g}\,\left[R-2\Lambda+ G_{\mu\nu} \, \g_{r_*}(\B) \,  R^{\mu\nu} \right],\label{nlffg}
\ee
where $G_{\mu\nu}$ is the Einstein tensor \Eq{Eiten} and
\be\label{fofag}
\g_{r_*}(\B) :=  \frac{\rme^{-r_*\B}-1}{\Box}\,.
\ee
This model is dictated by the above program (i)--(v) and may be also regarded as a phenomenological non-local limit of M-theory \cite{CaMo2}. The role of the non-local operator $1/\B$ is to compensate the second-order derivatives in curvature invariants. Its definition is presented in appendix \ref{app1}. To date, the perturbative renormalizability of the theory with \Eq{fofag} has been proven only with the use of the resummed propagator \cite{TBM}, while infinities have not been tamed yet in the orthodox expansion with the bare propagator. Nevertheless, this theory encodes all the main features of those non-local quantum gravities that have been shown to be renormalizable and its dynamics is simpler to deal with.

Even without considering gravity and the quantum limit, there is a general conceptual issue usually characterizing non-local physics. Namely, the Cauchy problem can be ill defined or highly non-standard in non-local theories \cite{Lew33,Car36,PU,Pau53}. In fact, while there is a time-honored tradition on \emph{linear} differential equations with infinitely many derivatives that admit a fair mathematical treatment \cite{Lew33,Car36}, \emph{non-linear} non-local equations such as those appearing in non-local field theories are a very different and much trickier business. For any tensorial field $\vp(t,{\bf x})$, it entails an infinite number of initial conditions $\vp(t_{\rm i},{\bf x})$, $\dot\vp(t_{\rm i},{\bf x})$, $\ddot\vp(t_{\rm i},{\bf x})$, \dots, representing an infinite number of degrees of freedom. As the Taylor expansion of $\vp(t,{\bf x})$ around $t=0$ is given by the full set of initial conditions, specifying the Cauchy problem would be tantamount to knowing the solution itself, if analytic \cite{MoZ}. This makes it very difficult to find analytic solutions to the equations of motion, even on Minkowski spacetime. Fortunately, the exponential operator \Eq{serep0} is under much greater control than other non-local operators, since (at least for finite $n$) the diffusion-equation method is available to find analytic solutions \cite{roll,cuta2,cuta3,cuta4,MuNu3,cuta5,cuta6,cuta7} which are well defined when perturbative expansions are not \cite{cuta2}. The Cauchy problem can be rendered meaningful, both in the free theory \cite{Car36,PU,BK1} and in the presence of interactions \cite{cuta3}. Consider a real scalar field $\phi(x)$ dependent on spacetime coordinates $x=(t,{\bf x})$. According to the diffusion-equation method, one promotes $\phi(t,{\bf x})$ to a field $\Phi(r,t,{\bf x})$ living in an extended spacetime with a fictitious extra coordinate $r$. This field is assumed to obey the diffusion equation $(\B-\p_r)\Phi(r,t,{\bf x})=0$, implemented at the level of the $(D+1)$-dimensional action by introducing an auxiliary scalar field $\chi(r,x)$ (dynamically constrained to be $\chi=\B\Phi$). Since the diffusion equation is linear in $\Phi$ (and $\chi$, consequently), the Laplace--Beltrami operator $\B$ commutes with the diffusion operator $\p_r$ and exponential operators act as translations on the extra coordinate, $\rme^{s\B}\Phi(r,t,{\bf x})=\rme^{s\p_r}\Phi(r,t,{\bf x})=\Phi(r+s,t,{\bf x})$. One can then show that, from the point of view of spacetime coordinates, the $(D+1)$-dimensional system is fully localized and that the only initial conditions to be specified are $\Phi(r,t_{\rm i},{\bf x})$, $\dot\Phi(r,t_{\rm i},{\bf x})$, $\chi(r,t_{\rm i},{\bf x})$, $\dot\chi(r,t_{\rm i},{\bf x})$ \cite{cuta3}. The infinite number of initial conditions $\phi(t_{\rm i},{\bf x})$, $\dot\phi(t_{\rm i},{\bf x})$, $\ddot\phi(t_{\rm i},{\bf x})$, \dots have been transferred into two initial conditions, which are actually boundary conditions in $r$, for an auxiliary field. When interactions are turned off, $\chi$ vanishes and one obtains the single degree of freedom, represented by $\phi(t_{\rm i},{\bf x})$ and $\dot\phi(t_{\rm i},{\bf x})$, of the free local theory.\footnote{This is obvious when integrating by parts the kinetic term, $\phi f(\B) \phi\to h(\B)\phi \B h(\B) \phi$, and absorbing non-locality with the field redefinition $\tilde\phi=h(\B)\phi$.} The original system is recovered when $r$ acquires a specific, fixed value proportional to the scale $r_*$. This value depends on the solution and is determined by solving the localized equations at $r=\b r_*$, where $\b$ is a constant. The resulting solutions $\phi(x)=\Phi(\b r_*,x)$ are not exact in general but they satisfy the equations of motion to a very good level of approximation \cite{roll,cuta5,cuta7}. 

For non-local gravity, one would like to apply the same method to the metric itself or to curvature invariants $\cR(g)$, but this is not possible in a direct way. Calling $\cR(r,x)$ the curvature invariants of a putative localized theory, since the diffusion equation $(\B-\p_r)\cR(r,x)=0$ would be non-linear in the metric $g_{\mu\nu}$, one would have
\be\label{probl}
[\B(g),\p_r]\cR(g)\neq 0\,,
\ee
and one would be unable to trade non-local operators for shifts in the extra direction. Moreover, the diffusion method applies for exponential operators, while in the actual quantum-gravity action \Eq{nlffg} non-locality is more complicated.

In this paper, we address this problem. First, we will use a field redefinition (already employed in other non-local gravities, although not for \Eq{nlffg} \cite{BMS,BCKM}, and similar to those used in scalar-tensor theories and modified gravity models) to transfer all non-locality to an auxiliary field $\phi_{\mu\nu}$. Next, we impose the diffusion equation on $\phi_{\mu\nu}$: the linearity problem is thus immediately solved and one can proceed to localize the non-local system, count the initial conditions and identify the degrees of freedom, which are finite in number. From there, one can begin the study of the dynamical solutions of the classical Einstein equations, but this goes beyond the scope of the present work. Counting non-local degrees of freedom is a subject surrounded by a certain halo of mystery and confusion in the literature. To make it hopefully clearer, we will make a long due comparison of the counting procedure and of its outcome in the methods proposed to date: the one based on the diffusion equation and the delocalization approach by Tomboulis \cite{Tom15}.

\subsection{Plan of the paper}

In preparation for the study of non-local gravity, the diffusion-equation method is reviewed in section \ref{scala} for a scalar field. This example is very useful because it contains virtually all the main ingredients we will need to localize non-local gravity and rewrite it in a user-friendly way: localized action, auxiliary fields, slicing choice, matching of the non-local and localized equations of motion, counting of degrees of freedom, solution of the Cauchy problem, and so on. The non-local scalar is introduced in section \ref{scala1}, while the localization procedure is described in section \ref{ized}. The counting of initial conditions and degrees of freedom is carried out in section \ref{scala3}, where we find that this number is, respectively, 2 and 1 for the real non-local scalar with non-linear interactions. Section \ref{solu} reviews another practical use of the diffusion-equation method, the construction of analytic solutions of the equations of motion. In section \ref{deloc}, we compare the diffusion-equation method with the results obtained in other approaches, mainly the delocalization method by Tomboulis \cite{Tom15}. A generalization of the method to non-local operators $\exp H(\B)$ with polynomial exponents is proposed in section \ref{Hge}, while non-polynomial profiles $H(\B)$ require some extra input which is discussed in a companion paper \cite{CMN3}.

The non-local gravitational action \Eq{nlffg} is studied in section \ref{eoms}, where we find the background-independent covariant Einstein equations for \emph{any} form factor $\g(\B)$ and recast the system in terms of an auxiliary field. Contrary to other calculations in the literature \cite{BCKM,Kos13,CKMT}, we find the equations of motion for an exponential-type form factor \Eq{fofag} in terms of parametric integrals rather than from the series expansion of the non-local operators. This new form is crucial both to solve the initial-value problem and to find explicit solutions with the diffusion-equation method.

The localized system corresponding to the non-local gravitational action \Eq{nlffg} is introduced and discussed in section \ref{locnlg}. After defining the localized action in section \ref{loac}, we obtain the equations of motion in section \ref{loeom}, which agree with the non-local ones. The counting of initial conditions and degrees of freedom is done in section \ref{lodof}, where we find that they amount to, respectively, 4 and $D(D-2)$. Appendices contain several technical details and the full derivation of the equations of motion.

Therefore, although in sections \ref{ized} and \ref{locnlg} we will concentrate on the form factor \Eq{fofag} for which renormalization is likely but still under debate, our results with auxiliary fields (section \ref{aux}) will be valid for an arbitrary form factor, while in section \ref{Hge} and in \cite{CMN3} we will generalize the diffusion-equation approach to form factors associated with finite quantum theories.

\subsection{Summary of main equations and claims}

To orient the reader, we summarize here the key formul\ae:
\begin{itemize}
\item Scalar field theory.
	\begin{itemize}
	\item Non-local action: \Eq{fac}.
	\item Non-local equation of motion: \Eq{tpheom}.
	\item Localized action: \Eq{act}.
	\item Localized equations of motion: \Eq{eomchi2}, \Eq{eomP12}, \Eq{eomP22}.
	\item Constraints on localized dynamics: \Eq{rstcon1}, \Eq{rstcon2}.
	\item Number of field degrees of freedom: \Eq{dof1}.
	\item Number of initial conditions: \Eq{dofic}.
	\end{itemize}
\item Gravity.
	\begin{itemize}
	\item Non-local action: \Eq{nlffgb}.
	\item Non-local equations of motion: \Eq{EinEq1}.
	\item Non-local action with auxiliary field: \Eq{nlff3}.
	\item Non-local equations of motion with auxiliary field: \Eq{eomnl1}.
	\item Localized action: \Eq{gactcm}.
	\item Localized equations of motion: \Eq{difPg}, \Eq{chimnde}, \Eq{uff}, \Eq{lasto}.
	\item Constraints on localized dynamics: \Eq{slic}, \Eq{Fmu2}.
	\item Number of field degrees of freedom: \Eq{dof3}.
	\item Number of initial conditions: \Eq{dof2}.
	\end{itemize}
\end{itemize}


\section{Diffusion-equation method: scalar field}\label{scala}

Before considering gravity, it will be useful to illustrate the main philosophy beyond and advantages of the diffusion-equation method. To this purpose, we review its application to a classical scalar field theory \cite{cuta3}, expanding the discussion therein to cover all important points that will help us to understand the results for non-local gravitational theories. We present a simplified version of the scalar system, with no nested integrals, no free parameters in the diffusion equation, and fewer assumptions than in \cite{cuta3}. The original version of \cite{cuta3} can be found in appendix \ref{app2}. A comparison between the scalar and gravitational systems will be done in section \ref{loeom}.


\subsection{Non-local system: traditional approach and problems}\label{scala1}

Consider the scalar-field action in $D$-dimensional Minkowski spacetime (with signature $-,+,\cdots,+$)
\be\label{fac}
S_{\phi} = \int \rmd^D x\,\cL_\phi,\qquad \cL_\phi =\frac12\phi\B\rme^{-r_*\B}\phi-V(\phi),
\ee
where $r_*$ is a constant of mass dimension $[r_*]=-2$ and $V(\phi)$ is a potential. We chose the exponential operator as the simplest example where the diffusion method works, but we will relax this assumption later to include operators of the form $\exp H(\B)$ not contemplated in the original treatment in \cite{cuta3}. Applying the variational principle to $S_\phi$, the equation of motion is
\be\label{tpheom}
\B \rme^{-r_*\B}\phi-V'(\phi)=0,
\ee
where a prime denotes a derivative with respect to $\phi$. The action \Eq{fac} and the dynamical equation \Eq{tpheom} are a prototype of, respectively, a \emph{non-local system} and a \emph{non-local equation of motion}.

The initial-condition problem associated with \Eq{tpheom} suffers from the conceptual issues outlined in the introduction. Rather than repeating the same mantra again, we recast the Cauchy problem as a problem of representation of the non-local operator $\exp(-r_*\B)$. To find a solution of \Eq{tpheom}, one must first define the left-hand side. The most obvious way to represent the exponential is via its series,
\be\label{opse}
\rme^{-r_*\B}=\sum_{n=0}^{+\infty}\frac{(-r_*\B)^n}{n!}=1-r_*\B+\frac12 r_*^2\B^2+\dots\,.
\ee
To find solutions, one can use different strategies. One of the oldest and most disastrous is to truncate the non-local operator up to some finite order $n_{\rm max}$. In doing so, one introduces instabilities corresponding to the Ostrogradski modes of a higher-derivative theory which has little or nothing to do with the starting theory \cite{MoZ,cuta2}. Exact procedures such as the root method exist for linear equations of motion \cite{PU,EW1,BK1,Ver09} but they have the disadvantage of being applicable only to non-interacting systems. Another possibility is to choose a profile $\phi(x)$ and apply the operator \Eq{opse}, but the series does not converge in general \cite{cuta2}. This does not necessarily mean that the chosen profile is not a solution of the equations of motion. Rather, the series representation \Eq{opse} is ill defined for a portion of the space of solutions. Even in the case where an exact solution is found, however, this may be non-unique for a given set of initial conditions \cite{EW1,MoZ,Tom15}.


\subsection{Localized system}\label{ized}

\begin{figure}
\centering
\includegraphics[width=8.2cm]{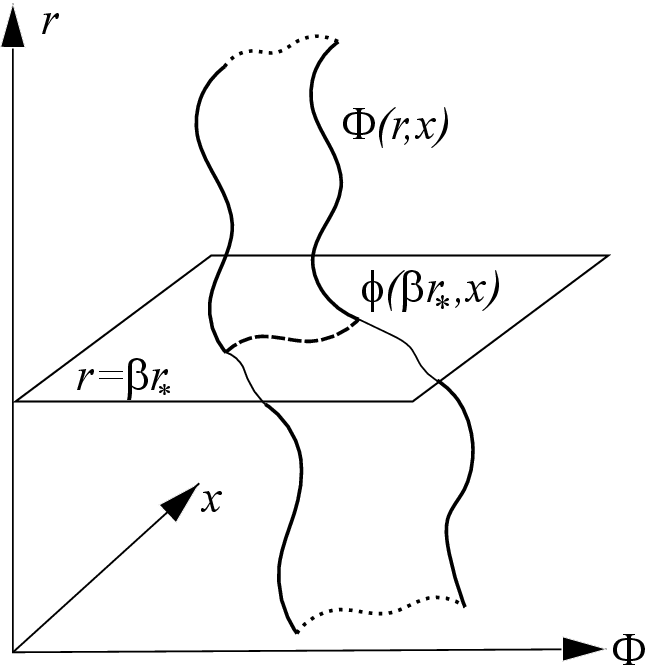}
\caption{\label{fig1} Diffusion-equation method describing the dynamics of the scalar field theory \Eq{fac} as the dynamics of the localized system \Eq{act} on the slice $r=\b r_*$.}
\end{figure}

The diffusion-equation method \cite{roll,cuta3,MuNu3,cuta7}, some elements of which can be found already in \cite{PU} (section III.B.3), bypasses the above-mentioned issues by converting the Cauchy problem into a boundary problem.\footnote{A similar attempt was made in \cite{CCG}.} All the non-locality is transferred into a fictitious extra direction $r$ and infinite initial conditions for the scalar field $\phi(t,{\bf x})$ are converted to a \emph{finite} number of field conditions on the $r=\b r_*$ slice along the extra direction, where $\b$ is a positive dimensionless constant (i.e., it is the physical value of $r$ measured in $r_*$ units). In other words, the rectangle $[0,\b r_*]\times [t_{\rm i},t_{\rm f}]$ can be spanned either along the $t$ (time) direction, as done when trying to solve the problem of initial conditions by brute force at $t=t_{\rm i}$, or along the $r$ direction, as done in the boundary-value problem with the diffusion method; see Fig.\ \ref{fig1} here and Fig.\ 1 of \cite{MuNu3}.

We will also be able to find the exact number of conditions required and to compare these results with those from other methods \cite{Tom15}.

\subsubsection{Lagrangian formalism}

The main idea is to exploit the fact that the exponential operator in \Eq{fac} acts as a translation operator if $\phi$ obeys a diffusion equation. Using this property, we can convert the non-local system into a \emph{localized} one where the diffusion equation is part of the dynamical equations, the field is evaluated at different points in an extra direction (along which the system is thus non-local), and only second-order derivative operators appear in the action and in the equations of motion. In this way, one can make sense of the Cauchy problem in the localized system and also in the non-local one, after establishing the conditions for which the two systems are equivalent \cite{cuta3}. This construction goes through some initial guesswork about the form of the correct localized system, especially regarding the integration domain of certain parts of the action, but this is not difficult in general. Both the scalar case \Eq{fac} and the gravitational action \Eq{nlffg} are simple enough to create no big trouble.

Let us therefore forget temporarily about the non-local system \Eq{fac} and consider the $(D+1)$-dimensional local system 
\ba
\cS[\Phi,\chi]&=&\int \rmd^D x\,\rmd r \left(\cL_{\Phi}+\cL_{\chi}\right)\,,\label{act}\\
\cL_{\Phi}&=&\frac12\Phi(r,x)\B \Phi(r-r_*,x)-V[\Phi(r,x)]\,,\label{locPh2}\\
\cL_{\chi}&=&\frac12 \int_0^{r_*} \rmd q\,\chi(r-q,x)(\p_{r'}-\B)\Phi(r',x)\,.\label{locch2}
\ea
where $r$ is an extra direction, $r_*$ is a specific value of $r$, $\Phi$ and $\chi$ are $(D+1)$-dimensional scalar fields and
\be\label{rprime}
r'=r+q-r_*\,,
\ee
hence $\p_{r'}=\p_q$. The action \Eq{act} is second-order (hence local) in spacetime derivatives and non-local in $r$ (because the fields take different arguments). The integration range of $r$ in \Eq{act} is arbitrary, it can be set equal to $r\in [0,+\infty)$ or any other interval containing  $[0,\b r_*]$ (the slices $r=0$ and $r=\b r_*$ play a special role: the former is the value where to specify the initial condition in $r$ of the diffusion equation, while the latter will be the physical value of the parameter $r$, for a given $\b$). 

The equations of motion are calculated from the infinitesimal variations of the action, using the functional derivative $\delta f(r,x)/\delta f(\bar r, \bar x)=\delta(r-\bar r)\delta^{(D)}(x-\bar x)$ for a field $f$. Since $\bar x$ and $\bar r$ are arbitrary, one can always assume the support of these delta distributions to lie within the integration domains in \Eq{act}, so that integrations in $x$, $r$ and $q$ are removed and the fields evaluated at $x=\bar x$ and $r=\bar r$. Bars will be removed in the final equations of motion.

The first variation we calculate is with respect to $\chi$. To keep notation light, let us ignore the trivially local $x$-dependence from now on. Doing it step by step,
\ba
0 &=& \frac{\de\cS[\Phi,\chi]}{\de\chi(\bar r,\bar x)}=\frac12 \int \rmd r \int_0^{r_*}\rmd q\,\de(r-q-\bar r)(\p_{r'}-\B)\Phi(r')\nonumber\\
&=& \frac12\int_{\bar r}^{r_*+\bar r} \rmd r (\p_{r'}-\B)\Phi(r')\Big|_{r'=2r-\bar r-r_*}.\label{inter2}
\ea
The integration of the Dirac distribution in $q$ gives the prescription $0<q=r-\bar r<r_*$, hence $\bar r<r< r_*+\bar r$ such that the support of the $\de$ lies in both $r$- and $q$-integration ranges. After a reparametrization $\rho= r-\bar r$, one gets
an integral of the form $\int_0^{r_*} \rmd \rho\, f(2\rho+\bar r-r_*)$. Since $\bar r$ is arbitrary, the integrand must be identically zero on shell for any integration range:
\be\label{eomchi2}
0=(\p_r-\B)\Phi(r,x)\,.
\ee
Another way to obtain the same result is to restrict from the very beginning the integration range in \Eq{act} from 0 to $+\infty$ or from $-\infty$ to $r_*$. In the first case, the integration range in \Eq{inter2} is reduced to $[0,r_*+\bar r]$, since $\bar r>0$. In the second case, the range in \Eq{inter2} is reduced to $[\bar r, r_*]$, since $\bar r<r_*$. In both cases, due to the arbitrariness of $\bar r$ the width of the integration domain is arbitrary, which implies that the integrand is zero.

\begin{figure}
\centering
\includegraphics[width=8.cm]{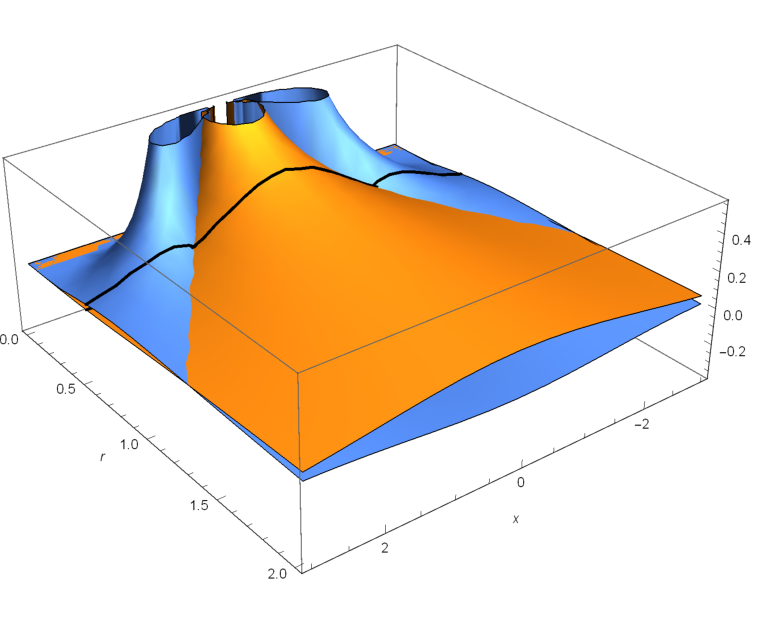}
\includegraphics[width=8.cm]{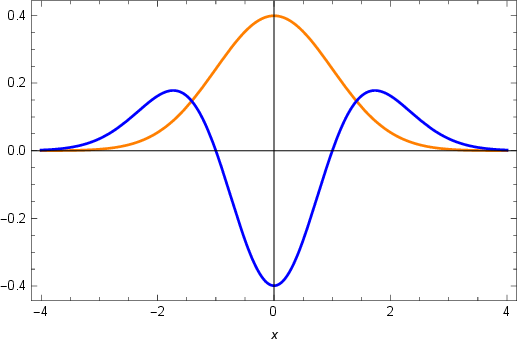}
\caption{\label{fig2} In $D=1$ flat Euclidean space, the solution of the diffusion equation \Eq{eomchi2} with initial condition $\Phi(0,x)=\de(x)$ is $\Phi(r,x)=\exp[-x^2/(4r)]/\sqrt{4\pi r}$. This solution is represented in the $(r,x)$ plane as an orange surface (concavity upwards) in the left plot, together with $\chi(r,x)=\p_x^2\Phi(r,x)$ (blue surface, concavity downwards). The section of these surfaces at $r=\b r_*=0.5$ (black thick line) are shown in the right plot.}
\end{figure}

The diffusion equation \Eq{eomchi2} is the first equation of motion. The second equation of motion is more complicated but very instructive, so that we report it in full. We integrate \Eq{locch2} by parts, in order to load all derivatives onto $\chi$:
\ba
\cL_{\chi}&=&\frac12\int_0^{r}\rmd q\,\p_{q}[\chi(r-q)\Phi(r')]-\frac12\int_0^{r}\rmd q\,\Phi(r')(\p_{q}+\B)\chi(r-q)\nonumber\\
&=&\frac12[\chi(r-r_*)\Phi(r)-\chi(r)\Phi(r-r_*)]-\frac12\int_0^{r_*} \rmd q\,\Phi(r')(\p_{r'}+\B)\chi(r-q)\,.\label{inpa2}
\ea
Therefore, varying with respect to $\Phi(\bar r,\bar x)$ gives
\ba
0=\frac{\de\cS[\Phi,\chi]}{\de\Phi(\bar r,\bar x)}&=&\frac12[\B\Phi(\bar r-r_*)+\chi(\bar r-r_*)]+\frac12[\B\Phi(\bar r+r_*)-\chi(\bar r+r_*)]-V'[\Phi(\bar r)]\nonumber\\
&&+\frac12\int_{\bar r}^{\bar r+r_*} \rmd r\,(\p_{-\bar r}-\B)\chi(2r-\bar r-r_*)\,.\label{inter22}
\ea
From this, we conclude that equation \Eq{inter22} reproduces \Eq{tpheom} if
\be\label{rstcon1}
\Phi(\b r_*,x)=\phi(x)\,,
\ee
where $\b>0$ is a real constant, and
\be\label{rstcon20}
\chi(r,x) = \B\Phi(r,x)\,.
\ee
In fact, in this case $\chi$ obeys the same diffusion equation \Eq{eomchi2} as $\Phi$, so that the two contributions in \Eq{inter22} must vanish separately, thus yielding the two equations of motion (restoring $x$-dependence)
\ba
\hspace{-.8cm} 0&=&\frac12[\B\Phi(r-r_*,x)+\chi(r-r_*,x)]+\frac12[\B\Phi(r+r_*)-\chi(r+r_*)]-V'[\Phi(r,x)]\,,\label{eomP12}\\
\hspace{-.8cm} 0&=&(\p_r-\B)\chi(r,x)\,.\label{eomP22}
\ea
Then, when evaluating \Eq{eomP12} at $r=\b r_*$ the first term yields $(1/2)2\B\rme^{-r_*\B}\Phi(\b r_*,x)=\B\rme^{-r_*\B}\phi(x)$, the second term vanishes and \Eq{eomP12} reproduces \Eq{tpheom} exactly. See Fig.\ \ref{fig2} for a toy example. Note that imposing \Eq{rstcon20} only at $r=\b r_*$,
\be\label{rstcon2}
\chi(\b r_*,x) = \B\Phi(\b r_*,x)\,,
\ee
or at any given $r=\tilde r$ instead of for all $r$ would again yield \Eq{rstcon20}, provided $\chi$ obeyed \Eq{eomP22}. In fact, parametrizing with $\s=r-\tilde r$, $\chi(r,x)= \chi(\s+ \tilde r,x) = \rme^{\s\B}\chi(\tilde r,x)= \rme^{\s\B}\B\Phi(\tilde r,x)=\B\rme^{\s\B}\Phi(\tilde r,x)= \B\Phi(\s+\tilde r,x)=\B\Phi(r,x)$. 

The introduction of the parameter $\b$ in \Eq{rstcon1} reflects the fact that the choice of the slice where the $(D+1)$-dimensional scalar field coincides with the $D$-dimensional field does not affect the final result. For instance, one could have chosen $\b=0$ and identified $\Phi(0,x)=\phi(x)$ (the ``initial'' condition in $r$ of the diffusion equation), $\chi(0,x)=\B\phi(x)$. However, in section \ref{solu} we will argue that equation \Eq{rstcon1} is far better suited than $\Phi(0,x)=\phi(x)$ for the task of finding dynamical solutions. This is why we introduced a strictly positive $\beta$ in the first place.

To summarize the logic here, given the non-local system \Eq{fac} one can always write down the system \Eq{act}--\Eq{locch2} localizing it. This localized system is not in one-to-one correspondence with the non-local system but it always admits, among its solutions, the solutions of the non-local system. These solutions are defined by the boundary condition \Eq{rstcon1} together with the local condition \Eq{rstcon2}. The sub-set of solutions of the localized system obeying these conditions are solutions to the original non-local one, since the above conditions are valid on shell (i.e., applying \Eq{eomchi2} and \Eq{eomP22} to \Eq{eomP12}). In other words, \Eq{rstcon1} and \Eq{rstcon2} define the sub-set of solutions of the localized system that recover the equations of motion and solutions of the original non-local system. Recalling that the localized system \Eq{act} must be reducible to the non-local one \Eq{fac} only at a certain slice $r=\b r_*$ in the extra direction, it is clear that we do not need to study the most general $(D+1)$-dimensional evolution of the localized dynamics, which is obtained by dropping \Eq{rstcon2}. 

Notice that it is not possible, while keeping the diffusing structure unaltered, to change the status of \Eq{rstcon2} from a condition imposed by hand to a consequence of the dynamics. For instance, one could try to add an extra term $\cL_\la=\la(r,x)[\B\Phi(r,x)-\chi(r,x)]$ to the action \Eq{act}, which would give \Eq{rstcon2} when varying $\cS[\Phi,\chi,\la]$ with respect to the Lagrange multiplier $\la$. However, equations \Eq{inter2} and \Eq{inter22} would become, respectively, $(\dots)-\lambda=0$ and $(\dots)+\B\lambda=0$, where the extra terms would vanish separately if, again, we imposed by hand
\be\label{lamb}
\la(r,x)=0\,.
\ee
This condition, replacing \Eq{rstcon2}, amounts to forbid source terms in the diffusion equation \Eq{eomchi2}. Indeed, the infinitely many degrees of freedom of the original non-local system are encoded in equation \Eq{rstcon2} or in the alternative equation \Eq{lamb}, both of which are a condition on the \emph{infinitely many} $r$-values of the fields $\Phi$ and $\chi$. Thus, demanding to get a fully self-determined diffusing localized system equivalent to the non-local one is not only impossible,\footnote{Of course, this claim does not apply to an arbitrary localized system not diffusing with the standard sourceless diffusion equation of Brownian motion.} but also meaningless, since the equivalence between the localized and the non-local system on one hand and the statement of the initial-value problem for the non-local system on the other hand must both go through the setting of an infinite number of conditions external to the dynamics.

For future use, we highlight three important features of the localization procedure which will apply, in their essence, also to the non-local gravity action \Eq{nlffg}.
\begin{enumerate}
\item By the diffusion-equation method, one does not establish a one-to-one correspondence between the localized system \Eq{act} and the non-local system \Eq{fac}. Rather, we showed that there exist field conditions on the $r=\b r_*$ slice such that the localized system has the same spacetime dynamics as the non-local system. This correspondence on a slice is depicted in Fig.\ \ref{fig1}.
\item To get the correct result, it was crucial to make a careful choice of the arguments in the diffusion-equation term \Eq{locch2} and a careful treatment of the boundary terms when integrating \Eq{locch2} by parts as in \Eq{inpa2}. Without such boundary terms, \Eq{eomP12} would have been unable to reproduce \Eq{tpheom} on the $r=\b r_*$ slice with the correct numerical factors.
\item The localized system is second-order in spacetime derivatives, for both $\Phi$ and $\chi$. Therefore, the Cauchy problem for this system, when restricted to spacetime directions $x^\mu$, is solved by \emph{four initial conditions} at some $t=t_{\rm i}$:
\be
\Phi(r,t_{\rm i},{\bf x}),\,\dot\Phi(r,t_{\rm i},{\bf x}),\,\chi(r,t_{\rm i},{\bf x}),\,\dot\chi(r,t_{\rm i},{\bf x})\,.
\ee
In particular, these conditions are valid at $r=\b r_*$, where, however, $\chi$ is fully determined once $\phi$ is known. Therefore, the Cauchy problem of the non-local system \Eq{fac} is solved by two initial conditions, corresponding (via \Eq{rstcon2}) to $\phi(t_{\rm i},{\bf x})$ and its first time derivative. 
\end{enumerate}

\subsubsection{Ghost mode}

In this subsection, we analyze a hidden ghost mode which, however, does not influence the non-local dynamics. To understand this aspect, we will employ a reformulation of the localized dynamics (equation \Eq{tildeL}), physically equivalent to \Eq{act}--\Eq{locch2}, which is convenient to study the degrees of freedom of the theory but is unsuitable for the practical treatment (Cauchy problem, solutions, and so on) of the dynamics, due to problems we will comment on in due course.

It is very well known that the kinetic term in \Eq{fac} can be symmetrized after integrating by part, so that the Lagrangian becomes
\be\label{facbis}
\cL_\phi =\frac12(\rme^{-\frac12r_*\B}\phi)\B(\rme^{-\frac12r_*\B}\phi)-V(\phi).
\ee
From here, one can make the field redefinition $\tilde\phi=\rme^{-\frac12r_*\B}\phi$ so often used in $p$-adic and string field theory. We will do something similar by considering the localized version of \Eq{facbis}, which is given by \Eq{act} with ($x$-dependence omitted everywhere)
\ba
\tilde\cL_{\Phi}=\frac12\Phi\left(r-\tfrac r_*\right)\B \Phi\left(r-\tfrac r_*\right)-V[\Phi(r,x)]\label{locPh2bis}
\ea
replacing \Eq{locPh2}. We note that the integral in \Eq{locch2} is pleonastic for the Laplace--Beltrami term, since both $\chi$ and $\Phi$ obey the diffusion equation:
\ba
\int\rmd^Dx\int_0^{r_*} \rmd q\,\chi(r-q)\B\Phi(r')&=&\int\rmd^Dx\int_0^{r_*} \rmd q\,\rme^{(\frac12r_*-q)\B}\chi\left(r-\tfrac r_*\right)\rme^{(q-\frac12r_*)\B}\B\Phi\left(r-\tfrac r_*\right)\nonumber\\
&=&\int\rmd^Dx\int_0^{r_*} \rmd q\,\chi\left(r-\tfrac r_*\right)\B\Phi\left(r-\tfrac r_*\right)\nonumber\\
&=&r_*\int\rmd^Dx\,\chi\left(r-\tfrac r_*\right)\B\Phi\left(r-\tfrac r_*\right)\,.
\ea
However, replacing \Eq{locch2} with a mixed term
\be\nonumber
\tilde\cL_{\chi}=-\frac{r_*}{2}\chi\left(r-\tfrac r_*\right)\B\Phi\left(r-\tfrac r_*\right)+\frac12 \int_0^{r_*} \rmd q\,\chi(r-q)\p_{r'}\Phi(r')
\ee
would not give the correct equations of motion, as we will see shortly. The reason is that $\tilde\cL_{\chi}$ is originated by an on-shell condition, a trick that invalidates the variational principle. To find the correct Lagrangian, we generalize this term with a generic functional of the fields, $\Phi\rightarrow f[\Phi,\chi]$. A last step we take (not necessary, but useful to simplify the physical interpretation) is to consider the field redefinitions
\be
\vp(r,x) :=\Phi\left(r-\tfrac r_*,x\right)-\frac{r_*}{2}\chi\left(r-\tfrac r_*,x\right),\qquad \psi(r,x):=\frac{r_*}{2}\chi\left(r-\tfrac r_*,x\right)\,,
\ee
so that
\be\label{Pvfpsi}
\Phi\left(r-\tfrac r_*\right)=\vp(r)+\psi(r)\,,
\ee
and the total Lagrangian on Minkowski spacetime is
\ba
\tilde\cL&=&-\frac12\p_\mu\Phi\left(r-\tfrac r_*\right)\p^\mu\Phi\left(r-\tfrac r_*\right)-V[\Phi(r)]+\frac{r_*}{2}\p_\mu\chi\left(r-\tfrac r_*\right)\p^\mu\Phi\left(r-\tfrac r_*\right)\nonumber\\
&&+\frac12 \int_0^{r_*} \rmd q\,\chi(r-q)\p_{r'}f(r')\nonumber\\
&=&-\frac12\p_\mu\vp(r)\p^\mu\vp(r)-V\left[\vp\left(r+\tfrac r_*\right)+\psi\left(r+\tfrac r_*\right)\right]+\frac12\p_\mu\psi(r)\p^\mu\psi(r)+\frac{1}{r_*}\,I(r)\,,\no \label{tildeL}
\ea
where
\ba\label{utile}
I(r) &:=& \int_{0}^{r_{*}}\rmd q\,\psi\left(r-q+\tfrac r_*\right)\p_q f\left(r+q-\tfrac r_*\right)\nonumber\\
&=& \psi\left(r-\tfrac r_*\right)f\left(r+\tfrac r_*\right)-\psi\left(r+\tfrac r_*\right)f\left(r-\tfrac r_*\right)\nonumber\\
&&-\int_0^{r_*}\rmd q\,f\left(r+q-\tfrac r_*\right)\p_q\psi\left(r-q+\tfrac r_*\right).
\ea
The function $f$ is determined in appendix \ref{appI} by requiring the recovery of the non-local dynamics on the $r=\b r_*$ slice.

Observing \Eq{tildeL}, one sees that the canonical scalar $\vp$ propagates with a kinetic term of the correct sign, while the canonical scalar $\psi$ (hence $\chi$) is a ghost. This detail went unnoticed in \cite{cuta3}.

There are two issues affecting \Eq{tildeL} and described in appendix \ref{appI}, but we should not lose sight of the reason why we introduced this Lagrangian. One may choose either \Eq{act}--\Eq{locch2} or \Eq{tildeL} depending on what one wants to study. For the analysis of the Cauchy problem and of the dynamical solutions, the action \Eq{act}--\Eq{locch2} is to be preferred, and in fact we will analyze non-local gravity under the same scheme. On the other hand, for the characterization (ghost-like or not) of the localized degrees of freedom the Lagrangian \Eq{tildeL}, or the Hamiltonian \Eq{hamil} we will derive from it in the next subsection, is more indicated. The counting of the localized degrees of freedom (section \ref{scala3}) can be performed indifferently in the original system \Eq{act}--\Eq{locch2}, in the Lagrangian \Eq{tildeL}, or in the Hamiltonian formalism derived from \Eq{tildeL}.

\subsubsection{Hamiltonian formalism}

To count the number of degrees of freedom in a non-local theory, we must first count the number of localized degrees of freedom in the associated localized $(D+1)$-dimensional theory. In the case of the scalar field, this information is already available in Lagrangian formalism, but for completeness we can obtain the same result from Hamiltonian formalism. The example presented in this subsection will illustrate the general method and its caveats. Its application in the localization of the scalar field was sketched in \cite{cuta3}, but here we will fill several gaps in that discussion. The actual counting of localized degrees of freedom will be done in section \ref{scala3}.

Although we do not write the non-local system \Eq{fac} in Hamiltonian formalism, we can reach a lesser but still instructive goal, namely, the formulation of the Hamiltonian approach for the associated localized system. However, if we take the localized system \Eq{act}--\Eq{locch2} as a starting point we soon meet several problems, all of which stem from the non-locality with respect to the $r$ direction. Momenta acquire a rather obscure non-invertible form and one cannot write down a Hamiltonian in phase space. However, the system is not constrained. We can avoid all the trouble by acting directly on \Eq{tildeL}. Calling $\tilde L := \int \rmd^{D-1}{\bf x}\int\rmd r\,\tilde\cL$ the Lagrangian, we can define the phase space and the Hamiltonian. The momenta are
\be
\pi_\vp(r,x) := \frac{\delta\tilde L}{\delta\dot\vp(r,x)}=\dot\vp(r,x),\qquad \pi_\psi(r,x) := \frac{\delta\tilde L}{\delta\dot\psi(r,x)}=-\dot\psi(r,x)\,.\label{pic}
\ee
Notice that, if we had calculated the momenta directly from \Eq{locPh2} and \Eq{locPh2bis}, we would have obtained $\pi_\Phi(r)=(1/2)[\dot\Phi(r-r_*)+\dot\Phi(r+r_*)-\int_0^{r_*}\rmd s\,\dot\chi(r-2s+r_*)]$ and $\pi_\chi(r) =-(1/2)\int_0^{r_*} \rmd s\,\dot\Phi(r+2s-r_*)$, which are not invertible locally with respect to $\dot\Phi(r)$ and $\dot\chi(r)$.

The non-vanishing equal-time Poisson brackets in terms of the spatial $(D-1)$-vectors ${\bf x}$ are
\ba
\{\vp(r_1,x_1),\,\pi_\vp(r_2,x_2)\}_{t_1=t_2} &=& \delta(r_1-r_2)\,\delta^{(D-1)}({\bf x}_1-{\bf x}_2)\,,\\
\{\psi(r_1,x_1),\,\pi_\psi(r_2,x_2)\}_{t_1=t_2} &=& \delta(r_1-r_2)\,\delta^{(D-1)}({\bf x}_1-{\bf x}_2)\,,
\ea
while the Hamiltonian of the system is ($x$-dependence omitted again)
\ba
H &:=&\int \rmd^{D-1}{\bf x} \rmd r \left[\pi_\vp(r)\dot\vp(r)+\pi_\psi(r)\dot\psi(r)\right]-\tilde L\nonumber\\
  &=& \int \rmd^{D-1}{\bf x} \rmd r \left\{\frac12\pi_\vp^2(r)+\frac12\N_i\vp(r)\N^i\vp(r)-\frac12\pi_\psi^2(r)-\frac12\N_i\psi(r)\N^i\psi(r)\right.\nonumber\\
	&&\qquad\qquad\qquad\left.+V[\Phi(r)]-\frac{1}{r_*} I[\psi(r),\Phi(r)]\right\}\,,\label{hamil}
\ea
where it is understood that $\Phi(r)=\vp(r+r_*/2)+\psi(r+r_*/2)$. Since $\Phi$ is shifted in $r$, $H$ is non-local in $r$ due to the terms in the last line of \Eq{hamil}. Nevertheless, the Hamiltonian is written solely in terms of phase-space variables and the phase-space fields are completely local in spacetime coordinates.

The evolution equations for the fields $\vp$ and $\psi$ trivially gives the momenta, $\dot\vp(r)=\{\vp(r),H\}=\de H/\de\pi_\vp(r)=\pi_\vp(r)$, $\dot\psi(r)=\{\psi(r),H\}=\de H/\de\pi_\psi(r)=-\pi_\psi(r)$, while the Hamiltonian evolution of the momenta give the localized equations of motion \Eq{eomH1} and \Eq{eomH2}:
\be
\dot\pi_\vp(\bar r) = \{\pi_\vp(\bar r),\,H\}=-\frac{\de H}{\de\vp(\bar r)}\,,\qquad \dot\pi_\psi(\bar r) = \{\pi_\psi(\bar r),\,H\}= -\frac{\de H}{\de\psi(\bar r)}\,.\label{eomH22}
\ee


\subsection{Initial conditions and degrees of freedom}\label{scala3}

The question about how many initial conditions we should specify for the non-local scalar system is related to another one: How many
degrees of freedom are hidden in equation \Eq{tpheom}? In higher-derivative theories, the presence of many degrees of freedom (Ostrogradski modes) is well known. For a system with $n$ derivatives, the Cauchy problem is uniquely solved by $n$ initial conditions. However, there is an uncrossable divide between higher-derivative and non-local theories, and one cannot conclude that non-local theories need $n=\infty$ initial conditions; conversely, truncating a non-local theory to finite order leads to a physically different model \cite{EW1,cuta2}.

To understand the problem, we review its root and also some confusion surrounding it. First of all, there is agreement in the literature about the fact that the \emph{free} system with constant, linear or quadratic $V(\phi)$
 has \emph{two} initial conditions. In the absence of interactions, the Cauchy problem associated with \Eq{tpheom} is specified only by $\phi(t_{\rm i},{\bf x})$ and $\dot\phi(t_{\rm i},{\bf x})$. The entire functional $\exp(-r_*\B)$ introduces no new poles in the spectrum of $\phi$ and the system is equivalent to the local one with $r_*=0$, as is obvious from the field redefinition $\tilde\phi=\rme^{-r_*\B/2}\phi$.\footnote{Another method, completely equivalent, is to work in Laplace momentum space.} This was first recognized as early as 1950 in the seminal paper by Pais and Uhlenbeck \cite{PU} (section III.B.3) and reiterated more recently, sometimes using very different terminology and techniques, in other works \cite{EW1,Sim90,BK1,cuta3}.

More contrived is the case with interactions. The reader unfamiliar with non-local theories may wonder why interactions should make any difference when counting the number of initial conditions. The reason is that, in this case, there is no field redefinition absorbing the non-local operator of the kinetic term. Any other rewriting will not work, either. For instance, a non-local kinetic term can always be expressed as a convolution with a kernel \cite{PU}. Consider the scalar-field Lagrangian $\cL_\phi=\phi f(\B)\,\phi-V(\phi)$ with generic form factor $f(\B)$. In momentum space, calling $F$ the Fourier transform of $f$,
\ba
\phi(x) f(\B)\,\phi(x) &=& \phi(x)\int\rmd^D k\,f(-k^2)\,\de(k^\mu-\rmi\N^\mu)\,\phi(x)\nonumber\\
											 &=& \phi(x)\int\rmd^D k\,\left[\int\frac{\rmd^D z}{(2\pi)^D}\,F(z)\,\rme^{-\rmi z^\mu k_\mu}\right]\,\de(k^\mu-\rmi\N^\mu)\,\phi(x)\nonumber\\
											 &=& \phi(x)\int\frac{\rmd^D z}{(2\pi)^D}\,F(z)\,\rme^{z^\mu\N_\mu}\phi(x)\nonumber\\
											 &=& \phi(x)\int\frac{\rmd^D z}{(2\pi)^D}\,F(z)\,\phi(x+z)\nonumber\\
											 &\stackrel{y:=z+x}{=}&\phi(x)\int\frac{\rmd^D y}{(2\pi)^D}\,F(y-x)\,\phi(y)\,.\label{ker}
\ea
Specifying the form factor $f(-k^2)$ determines the spectrum of the field. In general, the poles of the propagator correspond to the zeros of $f(-k^2)$ and to the poles of $F(z)$. This correspondence is straightforward for a massless dispersion relation $f(-k^2)=-k^{2n}$, where $F(z)\propto\de^{(2n)}(z)$ and $(2n)$ denotes the derivative of order $2n$ of the delta. The derivative order of the delta is the order of the pole. Polynomial dispersion relations have a similar structure, e.g., $f(-k^2)=-k^2-a k^{2n}$ gives $F(z)\propto \de^{(2)}(z)+a\de^{(2n)}(z)$. For $n=2$, this dispersion relation corresponds to one massive and one massless scalar mode, for a total of two double poles.\footnote{In fact, $f^{-1}(-k^2)=-[k^2(1+ak^2)]^{-1}=-k^{-2}+(a^{-1}+k^2)^{-1}$. The second mode is a ghost (positive residue).} Furthermore, when $f(\B)$ is non-local the field spectrum depends on whether the form factor is entire or not. In the case of \Eq{tpheom}, the propagator 
\be\label{prof}
f^{-1}(-k^2)=-\frac{\rme^{-r_*k^2}}{k^2} 
\ee
has a massless double pole, while $F(z)\propto (2r_*+z^2)\exp[z^2/(4r_*)]$ has a double massive pole. In the last two cases, the order and nature (massless or massive) of the particle poles and the poles of $F$ is less transparent, although their counting agrees.

From this exercise, it should become clear that hiding infinitely many derivatives into integrals with non-trivial kernels such as \Eq{ker}, or to transfer part of these derivatives onto the scalar potential and then converting them into integral operators, does not help in solving the Cauchy problem, since the two formulations are equivalent (on the space of real analytic functions \cite{MoZ}). In \cite{CMN3}, we complement this no-go result with its way out: If the kernel $F$ can be found by solving some finite-order differential equation extra with respect to the dynamical equations, then its contribution to the Cauchy problem becomes under full control.

The novelty brought in by the diffusion-equation method is that it allows one to go beyond the free theory and count the extra number of initial conditions. Surprisingly, in the scalar-field case this number is zero and there are no extra initial conditions with respect to the free theory. 

We reach this conclusion in three steps: (i) counting the number of field degrees of freedom of the localized theory; (ii) specifying the number of initial conditions (in time) for each localized field; (iii) restricting our attention to the slice $r=\b r_*$ where the non-local dynamics is recovered, and proceeding with the counting thereon. In Lagrangian formalism, we saw that there are two independent localized fields, either the pair $\Phi$ and $\chi$ or the pair $\vp$ and $\psi$. Consistently, the same result is obtained in Hamiltonian formalism, where there are two non-vanishing independent momenta $\pi_\vp$ and $\pi_\psi$. Since the dynamics is second-order in spacetime derivatives, there are two initial conditions per field, for a total of four.
\be\label{dof1}
\parbox[c]{13cm}{{\bf Number of degrees of freedom: scalar field.} \emph{The localized real scalar field theory \Eq{act}--\Eq{locch2} in $D+1$ dimensions has two scalar degrees of freedom $\Phi$ and $\chi$. On the $r$-slice where the system is equivalent to the non-local real scalar field theory \Eq{fac} in $D$ dimensions, the degree of freedom $\chi$ is no longer independent. Consequently, the non-local theory has \underline{one} non-perturbative scalar degree of freedom $\phi$.}}
\ee
\be\label{dofic}
\parbox[c]{13cm}{{\bf Number of initial conditions: scalar field.} \emph{The Cauchy problem on spacetime slices of the localized real scalar field theory \Eq{act}--\Eq{locch2} in $D+1$ dimensions is specified by four initial conditions $\Phi(r,t_{\rm i},{\bf x})$, $\dot\Phi(r,t_{\rm i},{\bf x})$, $\chi(r,t_{\rm i},{\bf x})$, $\dot\chi(r,t_{\rm i},{\bf x})$. As a consequence, the Cauchy problem of the non-local non-perturbative real scalar field theory \Eq{fac} in $D$ dimensions is specified by \underline{two} initial conditions $\phi(t_{\rm i},{\bf x})$ and $\dot\phi(t_{\rm i},{\bf x})$.}}
\ee

The nature of the new degree of freedom $\chi$ is quite peculiar. As we saw above with a diagonalization trick (used, for instance, also in \cite{DeWo2}), this field is a ghost and, in fact, the Hamiltonian \Eq{hamil} is unbounded from below.
 From the point of view of the $(D+1)$-dimensional localized system \Eq{act}--\Eq{locch2}, $\chi$ arises as a Lagrange multiplier introduced to enforce the diffusion equation of $\Phi$; $\chi$ itself does not appear in its own equation of motion \Eq{eomchi2}. Its $(D+1)$-dimensional dynamics, given by the equation of motion of $\Phi$, is non-trivial (in Hamiltonian formalism, the momentum $\pi_\chi\propto \pi_\psi$ does not vanish) but it only amounts to diffusion, equation \Eq{eomP22}. Eventually, it turned up that it is associated with $\Phi$ by the second-order derivative relation \Eq{rstcon2}. From the point of view of the $D$-dimensional non-local system, $\chi$ disappears because its diffusion is frozen at a given slice, and the dynamics is written solely in terms of $\phi$, its derivatives and its potential. At this point, there is only one degree of freedom whose \emph{perturbative classical} propagator \Eq{prof} describes a non-ghost massless scalar mode. The potentially dangerous ghost mode in the $(D+1)$-dimensional system turns out to be non-dynamical in $D$-dimensions and in the \emph{free} theory. 

In the interacting non-local theory, $\chi$ does play a part in the dynamics, but in the form of the potential for $\phi$. Combined with equation \Eq{tpheom}, the local condition \Eq{rstcon2} explains in part the finite proliferation of degrees of freedom in the interacting case. Since \Eq{rstcon2} implies $\chi(r,x)=\B\Phi(r,x)$ for all $r$, then from \Eq{tpheom} one has
\be\label{usefll}
\chi[(\b-1)r_*,x]=\B\Phi[(\b-1)r_*,x]=\B\rme^{-r_*\B}\Phi(\b r_*,x)=V'[\Phi(\b r_*,x)]=V'[\phi(x)]\,.
\ee
If $V\propto\phi^2$, then $\chi[(\b-1)r_*,x]\propto \Phi(\b r_*,x)=\phi(x)$ and there is no extra degree of freedom with respect to the $V=0$ case. For a cubic or higher-order polynomial, $\chi[(\b-1)r_*,x]$ is not linearly equivalent to $\phi$. Non-linearities can generate new degrees of freedom (a typical example is $f(R)$ gravity, which contains a hidden scalar mode apart from the graviton) but not in this case, since the field $\chi$ is not dynamical on the $r=\b r_*$ slice where \Eq{usefll} holds.


\subsection{Solutions}\label{solu}

Solutions of non-local theories can be categorized into perturbative and non-perturbative. Perturbative solutions can have two meanings, either as the solutions obtained when truncating the non-local operators to a finite order (a procedure we will not discuss here \cite{EW1,MoZ,cuta2}) or as the solutions obtained, order by order, starting from the free theory and modeling interactions as a perturbative series \cite{MoZ,EW1,JLM}. When all non-locality acts on interactions, the two meanings coincide. Non-perturbative solutions are all those solutions that cannot be reached in these ways and, in general, they constitute the great majority of all possible solutions of the system. The diffusion-equation method permits to get access precisely to these solutions with generic non-perturbative potential \cite{roll,cuta2,cuta4,cuta5,cuta6,cuta8}. 

When introducing the condition \Eq{rstcon1}, we commented on the fact that the identification of the localized dynamics with the non-local one could take place at any $r=\tilde r$ slice, including at $r=\tilde r=0$ where $\Phi(0,x)=\phi(x)$. However, for the sake of the construction of actual solutions this choice is not fortunate, since it corresponds to the initial condition of the heat kernel. In other words, setting the initial condition (in $r$) of the $(D+1)$-dimensional system to be the solution of the non-local system would take us back to the usual paradox with non-local dynamics, namely, that knowing all the infinite number of initial conditions (in time) $\phi(t_{\rm i},{\bf x}),\,\dot\phi(t_{\rm i},{\bf x}),\,\ddot\phi(t_{\rm i},{\bf x}),\,\dots$ is tantamount to already knowing the Taylor expansion of the full solution around $t=t_{\rm i}$. It is more logical, then, to impose \Eq{rstcon1} (the non-local solution is the outcome of the diffusion from $r=0$ to $r=\b r_*$ rather than of anti-diffusion from $r=\b r_*$ to $r=0$) and to set the initial condition $\Phi(0,x)$ in $r=0$ as something else. This ``something else'' can be most naturally recognized as the solution $\phi_{\rm loc}(x)$ of the \emph{local} system obtained by setting $r_*=0$ in equations \Eq{fac} and \Eq{tpheom}:
\be\label{filoc}
\Phi(0,x)=\phi_{\rm loc}(x)\,.
\ee
Then, the solution of the diffusion equation \Eq{eomchi2} can be found in integral form in momentum space. Calling $-k^2$ the eigenvalue of the Laplace--Beltrami operator $\B$ and writing
\be
\phi_{\rm loc}(x)=\int_{-\infty}^{+\infty}\frac{\rmd^D k}{(2\pi)^D}\,\rme^{-\rmi k\cdot x}\tilde\phi_{\rm loc}(k)\,,
\ee
one has
\be
\phi(x)=\Phi(\b r_*,x)=\rme^{\b r_*\B}\Phi(0,x)=\int_{-\infty}^{+\infty}\frac{\rmd^D k}{(2\pi)^D}\,\rme^{-\rmi k\cdot x}\rme^{-\b r_*k^2}\tilde\phi_{\rm loc}(k)\,.\label{soluz}
\ee
Since we know $\phi_{\rm loc}(x)$, we also know its Fourier transform $\tilde\phi_{\rm loc}(k)$ and we can obtain the full non-local solution $\phi(x)$. Examples of solutions of the scalar-field equation of motion \Eq{tpheom} using the diffusion-equation method can be found in \cite{cuta2} (on a Friedmann--Lema\^{i}tre--Robertson--Walker (FLRW) cosmological background), \cite{roll,cuta7} (Minkowski background, rolling tachyon of open string field theory), \cite{cuta3,cuta6} (Minkowski and FLRW backgrounds, $V\propto\Phi^n$ and $V\propto\exp(\la\Phi)$), \cite{cuta4,cuta5} (lump solutions on Minwkoski, FLRW and Euclidean backgrounds; kink solutions on Euclidean background), and \cite{cuta8} (FLRW solutions in a scalar-tensor non-local theory). Solutions of $p$-adic models, corresponding to \Eq{tpheom} without the $\B$ in the kinetic term, have been considered in \cite{cuta4,cuta7}. In some of these cases, a diffusion equation with opposite sign of the diffusion operator has been used, in which case the representation \Eq{soluz} may be ill-defined. This is not a problem, since there exist a more general integral form of the solution valid for any sign (see section 3.3 of \cite{cuta7}).

Note that, in general, convergence of the integral \Eq{soluz} will require $\b>0$. Also, setting $\b=1$ in \Eq{usefll} would yield $\B\phi_{\rm loc}=V'(\phi)$, implying $\phi_{\rm loc}=\phi$. To avoid this inconsistency, we exclude the value $\b=1$. Also, for any \emph{given} potential $V(\phi)$ and for a generic $0<\b<1$ the profile $\phi(x)$ is not a solution to the non-local equation of motion, not even approximately. Therefore, what one usually finds is an approximate solution $\phi(x)$ for a certain range of $x$. The actual value of $\b$ determines the limits of the $x$ range, since the profile typically depends on the combination $x^2/(4\b r_*)$. For instance, an approximated solution valid at large $x$ requires $\b>1$. However, cases are known where the profile $\phi(x)$ is an approximate solution for any $x$ (even small) with a very good degree of accuracy, which means that there exists a value of $\b$ such that the equation of motion is solved up to a maximal deviation of a few percent or less for some $x$, and with much greater accuracy everywhere else. These systems are related to (\cite{roll,cuta4,cuta5}) or inspired by (\cite{cuta2,cuta6}) string field theory. On the other hand, there may be special cases where $\phi$ is an exact solution, but these in general require a specifically tailored potential. Some examples of this inverse problem are given in \cite{cuta3}.


\subsection{Comparison with Tomboulis approach}\label{deloc}

Another approach handling non-perturbative solutions was proposed after the diffusion-equation method by Tomboulis \cite{Tom15}. Here, by a field redefinition one transfers non-local operators from the kinetic into the potential term, with a procedure analogous to that leading to \Eq{ker}. The latter is then written as an integral kernel, as above. This type of ``delocalized'' hyperbolic partial integro-differential equations are characterized by the phenomenon, due to the smearing of the kernel in \Eq{ker}, of ``spill-over'' (or delays) outside the standard causal cones of the local hyperbolic initial-value problem. Depending on the system, delays may be present only in the past or both in the past and in the future.

Comparing the diffusion-equation method with Tomboulis' delocalization (or delays) approach in classical theories, we find several similarities.
\begin{itemize}
\item Both recognize the central role of interactions to distinguish between local and non-local models.
\item Related to this, both agree also on the fact that, independently on whether one transfers non-locality from the kinetic term to interactions or not, it makes no sense to count the number of initial conditions just from the order of the kinetic term or by looking at any isolated part of the Lagrangian; in the limit of turning off the interactions, one may obtain the wrong answer. In this sense, the distinction between perturbative and non-perturbative solutions is not very useful in either method if one aims to make existence and uniqueness statements on the full dynamics.
\item The ill-defined concept of ``infinitely many initial conditions'' is traded with a boundary-value problem. In the diffusion-equation approach, the value of the $D$-dimensional field $\phi(t_{\rm i},{\bf x})$ and all its derivatives $\phi^{(n)}(t_{\rm i},{\bf x})$ at one time instant $t=t_{\rm i}$ is replaced by a field configuration $\Phi(r,t,{\bf x})$ living in $D+1$ dimensions, evaluated at a certain slice $r=\b r_*$ in the extra direction. In Tomboulis' approach, one specifies one or more functions rather than field values at one instant in the past, if delays occur only in the past light cone. If delays occur also in the future cone, as in systems with Lorentz-invariant interactions, then analogous specifications of functions must be done for them. For these systems, the type of non-local kernel has the same spill-over at all sides of the causal cone and the number of specifications in the future is finite and equal to the number of specifications in the past. Thus, both methods predict a \emph{finite, even number of conditions} (initial-value or boundary-value) for non-local scalar field theories. The specific prediction of the diffusion method is \Eq{dof1} and \Eq{dofic}, for any non-quadratic potential.
\item Consequently, because solutions are determined by picking conditions on $r$-slices in one case and past-future delay specifications in the other case, there are no implicit choices nor hidden conditions in the construction of such solutions, which are therefore unique once the explicit conditions are specified. This solves the long-standing problem of non-local theories where proving the existence of a solution by a brute-force \emph{Ansatz} does not imply, in the absence of any localization or delocalization method, its uniqueness \cite{EW1}.
\end{itemize}


\subsection{Generalizing to \texorpdfstring{$\exp H(\B)$}{} operators}\label{Hge}

Finally, let us comment on an extension of the above procedure to a non-locality of the form $\exp\B\to\exp H(\B)$ for some function $H$. In this case, one simply replaces $\B$ with $H(\B)$ in the Lagrange-multiplier equation \Eq{locch2}. Everything else follows suit. However, the system \Eq{act} is local and the Cauchy problem is well defined only if $H(\B)$ is a polynomial in the Laplace--Beltrami operator, $H(\B)=\sum_{n=1}^N a_n\B^n$. In this case, the number of initial conditions increases from 4 to $4n$: the value of the scalars $\Phi$ and $\chi$ at the initial time plus their first $2n-1$ derivatives.

For entire functions $H(\B)$, the ``localized'' system would be non-local. It may still be possible to localize \Eq{act} for special cases, for instance if $H(\B)=\exp\B$. However, in the most general case the diffusion method is insufficient to deal with these non-localities different from a pure exponential, unless an extra convolution equation is added to the system \cite{CMN3}.


\section{Non-local gravity: equations of motion}\label{eoms}

Consider the gravitational action
\be
\boxd{S_g = \frac{1}{2\kappa^2}\int \rmd^D x \sqrt{-g}\,\left[R-2\Lambda+G_{\mu\nu} \, \g(\B) \,  R^{\mu\nu} \right],\label{nlffgb}}
\ee
where $\g(\B)$ is a completely arbitrary form factor. In this section, we determine its dynamics in two ways. First, by a brute-force calculation, eventually specializable to the form factor \Eq{fofag}. Second, by recasting the system in terms of an auxiliary field.


\subsection{Einstein equations: pure gravity}

To compute the Einstein equations for a generic form factor $\g(\B)$, one must expand the latter in series of the Laplace--Beltrami operator $\B$,
\be\label{ggen}
\g=\sum_{n=0}^{+\infty}c_n\B^n\,,
\ee
where $c_n$ are constants, and vary with respect to the metric. We couple \Eq{nlffg} to matter minimally. Varying the total action $S=S_g+S_{\rm m}$ with respect to the contravariant metric $g^{\mu\nu}$, the matter part is dispensed with by the usual definition of energy-momentum tensor
\be\label{emt}
T_{\mu\nu} :=-\frac{2}{\sqrt{-g}}\frac{\delta S_{\rm m}}{\delta g^{\mu\nu}}\,.
\ee
In general, also matter fields will be non-local, but we do not consider their details here. The variations of curvature invariants and form factors are reported in appendix \ref{app3} and the full derivation of the final result is given in appendix \ref{app4}:
\ba
\k^2 T_{\mu\nu} &=& (1+\g\B) G_{\mu\nu}+\Lambda g_{\mu\nu}-\frac{1}{2} g_{\mu\nu}\,G_{\s\t}\g R^{\s\t}+2G^\s_{\ (\mu} \g  G_{\nu)\s}+g_{\mu\nu}\N^\s\N^\t\g G_{\s\t}\nonumber\\
&& -2\N^\s\N_{(\mu}\g G_{\nu)\s}+\frac12(G_{\mu\nu} \g R+R\g G_{\mu\nu})+\Theta_{\mu\nu}(R_{\s\t},G^{\s\t})\,,\label{EinEq1}
\ea
where the expression of $\Theta_{\mu\nu}(R_{\s\t},G^{\s\t})$ is given by \Eq{usef5a} for any form factor $\g$. The non-local equation \Eq{EinEq1} can be compared with similar ones found elsewhere \cite{BCKM,Kos13}.

For the particular choice of form factor \Eq{fofag},
\ba
(1+\g\B) G_{\mu\nu}&=&\rme^{-r_*\B}G_{\mu\nu}\,,\\
\Theta_{\mu\nu}(R_{\s\t},G^{\s\t}) &=&-\int_0^{r_*}\rmd q\,\bar\Theta_{\mu\nu}[\rme^{-q\B} R_{\s\t},\g_{r_*-q}(\B)G^{\s\t}]\,,\label{Theta2}
\ea
where $\bar\Theta_{\mu\nu}$ is given by equation \Eq{Theta}. The last expression, derived in appendix \ref{app4}, is fully explicit.


\subsection{Einstein equations: auxiliary field}\label{aux}

An alternative form of the Einstein equations makes use of an auxiliary field \cite{BMS,BCKM}. Here, we apply this method to \Eq{nlffg} for the first time. Consider the action
\be
\tilde S[g,\phi] = \frac{1}{2\kappa^2}\int \rmd^D x \sqrt{-g}\,\left[R-2\Lambda-2\phi^{\mu\nu} f_1(\B)\, R_{\mu\nu}+\left(\phi^{\mu\nu}-\frac{1}{D-2} g^{\mu\nu}\phi\right)\,f_2(\B)\,\phi_{\mu\nu}\right],\label{nlff2}
\ee
where $\phi_{\mu\nu}$ is a symmetric two-tensor, $\phi=\phi_\s^{\ \s}=g^{\mu\nu}\phi_{\mu\nu}$ is its trace and $f_{1,2}$ are some arbitrary form factors. The equations of motion for $\phi_{\mu\nu}$ are given by the variation $\de\tilde S/\de\phi^{\mu\nu}=0$:
\be\label{phieom}
-f_1(\B)\, R_{\mu\nu}+f_2(\B)\,\phi_{\mu\nu}-\frac{1}{D-2} g_{\mu\nu}\,f_2(\B)\,\phi=0\,.
\ee
Taking the trace of \Eq{phieom}, plugging it back and inverting for $\phi_{\mu\nu}$, one sees that
\be\label{bG}
\phi_{\mu\nu}=[f_2^{-1}f_1](\B)\,G_{\mu\nu}+\la_{\mu\nu}\,,\qquad \phi=-\left(\frac{D}{2}-1\right)[f_2^{-1}f_1](\B)\,R+\la_\mu^{\ \mu}\,,
\ee
where $\la_{\mu\nu}$ is the homogeneous solution of $f_2(\B)\la_{\mu\nu}=0$. Using \Eq{bG} and $f_2(\B)\la_{\mu\nu}=0$ in \Eq{nlff2} and integrating by parts, one gets the Lagrangian
\[
2\k^2\tilde\cL =R-2\Lambda-G_{\mu\nu}f_1f_2^{-1}f_1 R^{\mu\nu}-\la_{\mu\nu} f_1 R^{\mu\nu}\,.
\]
Comparing this with \Eq{nlffg}, we conclude that $\tilde S=S_g$ on shell provided
\be\label{ffg}
[f_1f_2^{-1}f_1](\B)=-\g(\B)\,,\qquad \la_{\mu\nu}=0\,.
\ee
There are various possible choices for the form factors $f_1$ and $f_2$; physically they are all equivalent as long as \Eq{ffg} holds. The simplest choice
\be
f_2=f_1=-\g
\ee
for an arbitrary form factor \Eq{ggen} satisfies the first condition in \Eq{ffg}, while only form factors with $c_0\neq 0$ (i.e., those with trivial kernel) also guarantee that the second condition in \Eq{ffg} is obeyed. In fact, $\g(\B)=c_0+c_1\B+O(\B^2)$, so that $\g(\B)\la_{\mu\nu}=0$ if, and only if, $\la_{\mu\nu}\equiv 0$. The form factor \Eq{fofag} is of this type, since $\g_{r_*}(\B)=-r_*+(r_*^2/2)\B+O(\B^2)$. In other words, there is no homogeneous solution we should worry about when recasting \Eq{nlffg} as \Eq{nlff2}, contrary to what happens when making field redefinitions in $f(\B^{-1}R)$ non-local gravity \cite{NoOd,Kos08,MaMa,ZKSZ}.

Thus, \Eq{nlff2} becomes
\be
\boxd{\tilde S[g,\phi] = \frac{1}{2\kappa^2}\int \rmd^D x \sqrt{-g}\,\left[R-2\Lambda+\left(2R_{\mu\nu}-\phi_{\mu\nu}+\frac{1}{D-2} g_{\mu\nu}\phi\right)\,\g(\B)\,\phi^{\mu\nu}\right].\label{nlff3}}
\ee
In appendix \ref{app5}, we show that the covariant equations of motion of the theory \Eq{nlff3} are
\ba
\k^2 T_{\mu\nu} &=& G_{\mu\nu}+\B\g \phi_{\mu\nu}+\Lambda g_{\mu\nu}-\frac12 g_{\mu\nu} X_{\s\t}\g\phi^{\s\t}+2\phi_{(\mu}^{\ \s}\g\phi_{\nu)\s}+g_{\mu\nu}\N^\s\N^\t \g\phi_{\s\t}\nonumber\\
								&& -2\N^\s\N_{(\mu} \g\phi_{\nu)\s}-\frac{1}{D-2}(\phi_{\mu\nu}\g\phi+\phi\g\phi_{\mu\nu})+\Theta_{\mu\nu}(X_{\s\t},\phi^{\s\t})\,,\label{eomnl1}\\
X_{\s\t}&:=&2R_{\s\t}-\phi_{\s\t}+\frac{1}{D-2}g_{\s\t}\phi\,,\label{X}
\ea
accompanied by the equation of motion $\de \tilde S[g,\phi]/\de\phi^{\mu\nu}=0$ and its trace:
\be\label{eomnl2}
\phi_{\mu\nu}=G_{\mu\nu}\qquad \Rightarrow \qquad \phi=G=-\frac{D-2}{2}R\,,\qquad X_{\mu\nu}=R_{\mu\nu}\,.
\ee
We call \Eq{eomnl1} Einstein equations because they come from the variation of the metric and \Eq{eomnl2} Einstein-like equations because they resemble the Einstein equations of general relativity, where $\phi_{\mu\nu}$ plays the role of a stress-energy tensor. Notice from \Eq{eomnl2} that the field $\phi_{\mu\nu}$ is local and does not hide $1/\B$ operators. This check \emph{a posteriori} guarantees that the ordinary variational principle (where fields and their first derivatives vanish at infinity) has been correctly applied.

Consistently, \Eq{EinEq1} and \Eq{eomnl1} agree on shell, i.e., when \Eq{eomnl2} is used (see appendix \ref{app6}).


\subsection{Brief remarks on causality}

Whenever a factor $\B^{-1}$ appears in a non-local theory, causality may be in trouble. The line of reasoning is well known and relies on the definition of the inverse d'Alembertian through the Green equation \Eq{bcK}, where one must specify a contour prescription for the Green function $\cK$. The main point is that even if the causal (retarded) propagator $\cK_{\rm ret}(x-y)$ is used to define the $\B^{-1}$ operator, a variation of the action with respect to the fundamental fields always gives rise to the even combination
\be\nonumber
\cK_{\rm ret}(x-y)+\cK_{\rm ret}(y-x)=:\cK_{\rm ret}(x-y)+ \cK_{\rm adv}(x-y)\,.
\ee
The retarded Green function is not even, and changing sign to its argument gives the advanced Green function $\cK_{\rm ret}(y-x) = \cK_{\rm adv}(x-y)$, which is anti-causal. Therefore, the equations of motion obtained from theories with non-localities of the type $\B^{-1}$ (typically, theories where the quantum effective action, not the classical one, is non-local) are necessarily acausal \cite{BDFM}.

However, this argument does not apply in our case because the non-localities we deal with do not need any prescription for the $\B^{-1}$ factor, as it always appears in a combination $\g(\B)=c_0+c_1\B+O(\B^2)$ which is analytic when ``$\B =0$'' (in particular, $\g_{r_*}(\B)=-r_*+(r_*^2/2)\B+O(\B^2)$). In other words, in all the theories of quantum gravity with a fundamental non-locality, non-localities (at the level of the classical action, not of the quantum effective one) are always of the type \Eq{effeb} with $f(0)=0$. As a consequence of this fact, the Green function associated with the non-local operator $\g(\B)$ is symmetric.

A one-dimensional example in flat space will further clarify the matter. The non-local operator containing $\B^{-1}$ is (\ref{fofag}), which can be written as \Eq{fofag2}. Its Green function $K(x-y)$ is the solution of the equation
\be\label{exerspace}
\g_{r_*}(\B_x)K(x-y)=-\int_{0}^{r_*} \rmd s\, \rme^{- s\B_x} \, K(x-y)= \delta(x-y)\,,
\ee
or, in momentum space, 
\be\label{exermom}
\int_{0}^{r_*} \rmd s\, \rme^{s k^2} \, \tilde K(k)=-1\,.
\ee
While the inverse of the $\B$ operator needs to be prescribed because the naive solution $1/k^2$ of the Green equation does not define a tempered distribution, the solution of eq.\ (\ref{exermom}) does not need to be regularized, as its algebraic solution
\be \label{exersolk}
\tilde K (k) = - \left[ \int_{0}^{r_*} \rmd s\,\rme^{s k^2} \right]^{-1} = - \frac{k^2}{\rme^{r_* k^2}-1}
\ee
already defines a tempered distribution. Its Fourier transform cannot be written in closed form but is very well behaved and, most importantly, is manifestly symmetric under the exchange $x\leftrightarrow  y$,\footnote{In the $\B^{-1}$ case, the regularization procedure needed to define $1/k^2$ as a tempered distribution prevents the Green function to be symmetric under the exchange $x \leftrightarrow y$, leading to the known mismatch between causality and symmetry of the propagator \cite{BDFM}.}
\be
\label{exersolx}
K(x-y)= -\frac{1}{\pi} \int_0^\infty\rmd k\, \frac{k^2\, \cos [k (x-y)]}{\rme^{r_* k^2}-1}\,.
\ee
	
The fact that the $\B^{-1}$ operator in the form factor $\g(\B)$ of fundamentally non-local quantum gravity does not introduce causality breaking is not, by itself, a guarantee of causality of these theories but, at least, it shows that standard arguments against causality, plaguing effective non-local field theories, do not apply in our case. The problem of causality in fundamentally non-local theories is subtle \cite{Tom15} and might not admit an all-or-nothing solution, in the sense that the theory might retain macrocausality \cite{GiMo} while including acceptable violations of microcausality. This interesting possibility will be explored elsewhere.


\section{Localization of non-local gravity}\label{locnlg}

In section \ref{eoms}, we started from a gravitational action $S_g[g(x)]$ and introduced an auxiliary tensor field $\phi_{\mu\nu}$ so that we could rewrite the original action as a functional of this field and the metric, $S_g[g(x)]= \tilde S[g_{\mu\nu}(x),\phi_{\mu\nu}(x)]$, where the gravitational part of $\tilde S$ is given by the integral of \Eq{nlff3}. In this section, we will construct a functional $\cS_g[g_{\mu\nu}(x),\Phi_{\mu\nu}(r,x),\chi_{\mu\nu}(r,x),$ $\la_{\mu\nu}(r,x)]$ representing a system living in $D+1$ dimensions and local in spacetime coordinates. Here we show that the two systems coincide at a section $r=\b r_*$ in the $(D+1)$-dimensional space,
\be\label{sss}
\cS_g[g_{\mu\nu}(\b r_*,x),\Phi_{\mu\nu}(\b r_*,x),\chi_{\mu\nu}(\b r_*,x)]=\tilde S_g[g_{\mu\nu}(x),\phi_{\mu\nu}(x),\chi_{\mu\nu}(x)]=S_g[g_{\mu\nu}(x)]\,,
\ee
where the equalities are meant to be valid on-shell, i.e., at the level of the dynamics. This statement, which can be immediately extended to actions that include also matter fields, is the extension to gravity of the results of section \ref{scala} \cite{cuta3} for a scalar field in Minkowski spacetime.


\subsection{Localized action}\label{loac}

We apply the procedure illustrated in section \ref{scala} to the gravitational theory with form factor \Eq{fofag}. We have seen that \Eq{nlffg} is physically equivalent to the action \Eq{nlff3}, which can also be written as
\be\label{phint}
\tilde S[g,\phi] = \frac{1}{2\kappa^2}\int \rmd^D x \sqrt{-g}\,\left[R-2\Lambda-\int_0^{r_*}\rmd s\left(2R_{\mu\nu}-\phi_{\mu\nu}+\frac{1}{D-2} g_{\mu\nu}\phi\right)\rme^{-s\B}\phi^{\mu\nu}\right]
\ee
thanks to \Eq{fofag2}. Using \Eq{phint} instead of \Eq{nlffg} will allow us to enforce the diffusion equation to $(D+1)$-dimensional fields without facing the commutation problem \Eq{probl} mentioned in the introduction and the fact that the metric field does not obey a linear diffusion equation. This problem is solved by letting only auxiliary fields diffuse, while the gravitational field does not diffuse at all: it is a dynamical field living in a fixed $r=\b r_*$ slice. Therefore, an interesting difference with respect to the scalar-field case is that here some fields (which we will call $\Phi_{\mu\nu}(r,x)$ and $\chi_{\mu\nu}(r,x)$) are free to evolve in the whole $(D+1)$-dimensional bulk, while others (the metric $g_{\mu\nu}(x)$ and the Ricci tensor $R_{\mu\nu}(x)$ derived from it) are confined into the slice where the higher-dimensional localized system is made equivalent to the non-local one. This configuration strongly reminds us of braneworld scenarios where $r$ is the direction transverse to a brane at $r=\b r_*$ and the Einstein--Hilbert Lagrangian contributes with a term $[R(x)-2\Lambda]\,\de(r-\b r_*)$. Another possibility, which we will follow from now on and yields the same result, is to consider an $r$-dependent $g_{\mu\nu}(r,x)$ dynamically constrained to be constant along $r$:
\be
\boxd{\cS_g=\frac{1}{2\k^2}\int\rmd^Dx\,\rmd r\,\sqrt{-g(r)}\left(\cL_R+\cL_\Phi+\cL_\chi+\cL_\la\right),\label{gactcm}}
\ee
\ba
\cL_R\!\! &=&\!\! R(r)-2\Lambda\,,\\
\cL_\Phi\!\! &=&\!\! -\int_0^{r_*}\rmd s\,\left[2\cR_{\mu\nu}(r)-\Phi_{\mu\nu}(r)+\frac{1}{D-2}g_{\mu\nu}(r)\Phi(r)\right]\Phi^{\mu\nu}(r-s)\,,\label{gact2cm}\\
\cL_\chi\!\! &=&\!\! -\int_0^{r_*}\rmd s\int_0^{s}\rmd q\,\chi_{\mu\nu}(r-q)(\p_{r'}-\B)\Phi^{\mu\nu}(r')\,,\label{difeqgcm}\\
\cL_\la\!\!  &=&\!\! \la_{\mu\nu}(r)\,\p_r g^{\mu\nu}(r)\,,\label{lagcm}
\ea
where the metric is $r$-dependent just like the other fields, its Ricci curvature is denoted with a curly $\cR_{\mu\nu}$, we introduced a Lagrange multiplier $\la_{\mu\nu}$, we omitted the $x$-dependence everywhere, $\Phi= g^{\mu\nu}\Phi_{\mu\nu}$ is the trace of the symmetric rank-2 tensor $\Phi_{\mu\nu}$, and
\be\label{rprime2}
r'=r+q-s\,,
\ee
so that $\p_{r'}=\p_q$ in \Eq{difeqgcm}. All tensorial indices still run from 0 to $D-1$, so that the theory \Eq{gactcm} is a fake $D+1$ system, which is not $(D+1)$-covariant anyway due to the diffusion equation term. In analogy with \Eq{X}, it will be convenient to define the tensorial combination
\be\label{Xr}
X_{\mu\nu}(r):=2\cR_{\mu\nu}(r)-\Phi_{\mu\nu}(r)+\frac{1}{D-2}g_{\mu\nu}(r)\Phi(r)\,.
\ee

Comparing with the scalar field theory \Eq{act}, there are four major differences one should note: (a) all fields are rank-2 tensors; (b) there is an extra integration $-\int_0^{r_*}\rmd s$ accounting for the more complicated form factor \Eq{fofag2}; (c) because of (b), the $q$-integral in \Eq{difeqgcm} is nested, while in \Eq{locch2} it is definite; (d) because of (c), \Eq{rprime2} replaces the scalar-field parameter \Eq{rprime}.


\subsection{Localized equations of motion}\label{loeom}

In intermediate steps of the derivation, we will omit the $x$-dependence in all fields as well as the discussions of section \ref{scala} on integration domains. The equation of motion for $\la_{\mu\nu}$ establishes the independence of the metric from the extra coordinate $r$:
\be\label{eomla}
0=\frac{\de\cS_g}{\de\la^{\mu\nu}(\bar r)}=\p_{\bar r} g_{\mu\nu}(\bar r,x)\qquad\Rightarrow\qquad g_{\mu\nu}(r,x)=g_{\mu\nu}(x)\,.
\ee
Therefore, in the following we can apply this equation on shell and ignore any change (shift, integration, and so on) in the $r$-argument of the metric, of the Laplace--Beltrami operator, and of curvature invariants, unless stated otherwise. The equations of motion turn out to be
\ba
0&=&(\p_r-\B)\Phi_{\mu\nu}(r,x)\,,\label{difPg}\\
0&=&(\p_r-\B)\chi_{\mu\nu}(r,x)\,,\label{chimnde}\\
0&=&\int_0^{r_*}\rmd s\left[X_{\mu\nu}(\bar r-s)+X_{\mu\nu}(\bar r+s)-2\cR_{\mu\nu}(r-s)+\chi_{\mu\nu}(\bar r-s)-\chi_{\mu\nu}(\bar r+s)\right]\,,\label{uff}\\
\k^2T_{\mu\nu} &=& G_{\mu\nu}+\Lambda g_{\mu\nu}-\int_0^{r_*}\rmd s\left\{\vphantom{\frac{1}{D-2}}-\frac12\,g_{\mu\nu}X_{\s\t}(r)\Phi^{\s\t}(r-s)+2\Phi_{\s(\mu}(r)\Phi_{\nu)}^{\ \ \s}(r-s)\right.\nonumber\\
&&+\B\Phi_{\mu\nu}(r-s)+g_{\mu\nu}\N^\s\N^\t\Phi_{\s\t}(r-s)-2\N^\s\N_{(\mu}\Phi_{\nu)\s}(r-s)\nonumber\\
&& \left.-\frac{1}{D-2}\left[\Phi_{\mu\nu}(r)\Phi(r-s)+\Phi(r)\Phi_{\mu\nu}(r-s)\right]-\int_0^{s}\rmd q\,\bar\Theta_{\mu\nu}[\chi_{\s\t}(r-q),\Phi^{\s\t}(r+q-s)]\right\}\!.\no \label{lasto}
\ea

Let us see where they come from. The equation of motion for $\chi_{\mu\nu}$ is
\ba
\hspace{-.8cm}0 &=& \frac{\de\cS_g}{\de\chi^{\mu\nu}(\bar r)}=-\int \rmd r \int_0^{r_*}\rmd s\int_0^{s}\rmd q\,\de(r-q-\bar r)(\p_{r'}-\B)\Phi_{\mu\nu}(r')\nonumber\\
  &\stackrel{\textrm{\tiny \Eq{eomla}}}{=}& -\int_{\bar r}^{\bar r+r_*} \rmd r\int_0^{r_*}\rmd s\,(\p_{r'}-\B)\Phi_{\mu\nu}(r')\Big|_{r'=2r-\bar r-s}\nonumber\\
  &=&-\int_0^{r_*}\rmd\rho\int_0^{r_*}\rmd s\,(\p_{r'}-\B)\Phi_{\mu\nu}(r')\Big|_{r'=2\rho+\bar r-s}\,,\label{inter2g}
\ea
where we first integrated in $q$, then restricted the integration in $r$ from the condition $0<q=r-\bar r<s<r_*$, and then made a reparametrization $\rho=r-\bar r$. Since $\bar r$ and hence $r'$ in the argument of the integrand is arbitrary, we get the diffusion equation \Eq{difPg}. Integrating first in $r$ or, after using \Eq{intint}, in $s$ would yield the same result. 

To obtain the equation of motion for $\Phi_{\mu\nu}$, we note that
\ba
\cL_{\chi}&\stackrel{\textrm{\tiny \Eq{eomla}}}{=}& \int_0^{r_*}\rmd s\int_0^{s}\rmd q\,\left\{-\p_q[\chi_{\mu\nu}(r-q)\Phi^{\mu\nu}(r')]+\Phi^{\mu\nu}(r')(\p_{r'}+\B)\chi_{\mu\nu}(r-q)\right\}\nonumber\\
&=&-\int_0^{r_*}\rmd s\left[\chi_{\mu\nu}(r-s)\Phi^{\mu\nu}(r)-\chi_{\mu\nu}(r)\Phi^{\mu\nu}(r-s)\right]\no 
&&+\int_0^{r_*}\rmd s\int_0^{s}\rmd q\,\Phi^{\mu\nu}(r')(\p_{r'}+\B)\chi_{\mu\nu}(r-q),\label{inpa2g}
\ea
where we used \Eq{eomla} to make $\sqrt{-g}$ slide through integration by parts (or, more precisely, we omitted terms that vanish on shell). This expression is the doubly integrated tensor-field analogue of \Eq{inpa2}. Then,
\ba
0&=&\frac{\de\cS_g}{\de\Phi^{\mu\nu}(\bar r)}\nonumber\\
&=&-\int_0^{r_*}\rmd s\left[X_{\mu\nu}(\bar r-s)+X_{\mu\nu}(\bar r+s)-2\cR_{\mu\nu}(r-s)+\chi_{\mu\nu}(\bar r-s)-\chi_{\mu\nu}(\bar r+s)\right]\nonumber\\
&&+\int_0^{r_*}\rmd s\int_0^{s+\bar r} \rmd r\,(\p_{\bar r}+\B)\chi_{\mu\nu}(2r -\bar r-s)\,.\label{inter22g}
\ea
Arbitrariness of $\bar r$ implies the diffusion equation \Eq{chimnde} and we are left with the equation of motion \Eq{uff}. We can easily find special solutions of this equation. For instance, it vanishes identically if we set the integrand to be zero. As in the scalar-field case, we can satisfy this equation by imposing certain constraints on the fields. The only one that respects $s$-shifts is
\be\label{Fmu}
\chi_{\mu\nu}(r,x)=X_{\mu\nu}(r,x)\stackrel{?}{=}\cR_{\mu\nu}(r,x)\,.
\ee
Therefore, just like in the scalar-field case, there is a condition imposed by hand such that one or more equations of motions are solved.\footnote{Other types of solution can be found by factorizing the $r$-dependence in $X_{\mu\nu}$ and $\chi_{\mu\nu}$. In fact, \Eq{uff} is of the form $\int_0^{r_*}\rmd s\,[A_{\mu\nu}(r+s;x)+B_{\mu\nu}(r-s;x)]=2R_{\mu\nu}(x)$. Writing $A_{\mu\nu}(r+s;x)=a(r+s)\,a_{\mu\nu}(x)$ and $B_{\mu\nu}(r-s;x)=b(r-s)\,b_{\mu\nu}(x)$ with $a_{\mu\nu}+b_{\mu\nu}=2R_{\mu\nu}$, the relation $\int_0^{r_*}\rmd s\,[a(r+s)+b(r-s)]=1$ can be solved by linear or trigonometric functions. The diffusion equations \Eq{difPg} and \Eq{chimnde} then become eigenvalue equations of the type $\B \Phi_{\mu\nu}\propto \Phi_{\mu\nu}$. These solutions, which were called ``stationary'' in \cite{cuta8} since they do not diffuse non-trivially, are still popular because they allow to solve the non-local cosmological equations of motion directly \cite{BKM1,KSKS}, although they are very limited tools when attempting to solve more general non-local systems \cite{cuta5,cuta7,cuta8}.} We recognize some similarities between the scalar and the gravity case. In the scalar example, \Eq{eomP22} held and one checked that \Eq{eomP12} reproduced \Eq{tpheom} provided equation \Eq{rstcon20}, $\chi(r,x) = \B\Phi(r,x)$, held. In the gravitational case, the analogue of the diffusion equation \Eq{eomP22} for $\chi$ is simply \Eq{chimnde} for $\chi_{\mu\nu}$, while the analogue of \Eq{rstcon20} is \Eq{Fmu}. 

The similarities do not stop here. In fact, both \Eq{rstcon20} and \Eq{Fmu} are second-order derivative relations, $\chi\sim \p^2\Phi$ and $\chi_{\mu\nu}\sim \p^2 g_{\mu\nu}$. Also, in the gravity case there are two auxiliary fields instead of one, $\Phi_{\mu\nu}(r,x)$ and $\chi_{\mu\nu}(r,x)$, but only the latter has been introduced exclusively in the context of the localized system, just like the scalar $\chi$ in section \ref{ized}. $\Phi_{\mu\nu}$ is ``auxiliary'' only because, on the slice $r=\b r_*$, it is equivalent to the field $\phi_{\mu\nu}$ introduced to recast the non-local gravitational theory in a convenient way:
\be\label{slic}
\Phi_{\mu\nu}(\b r_*,x)=\phi_{\mu\nu}(x)\,.
\ee

The last parallelism we can draw with the scalar theory is between the condition $\chi=\B\Phi$, imposed to match the equation of motion of the localized scalar field with the non-local equation of motion, and equation \Eq{Fmu}, which does exactly the same service albeit in a subtler way. In fact, \Eq{Fmu} is needed to satisfy one of the dynamical localized equations, but in retrospective it cannot be valid for all $r$ because of \Eq{eomla}: on shell, the rightmost member is independent of $r$. To put it in other words, we reached \Eq{Fmu} only because we took advantage of the pleonastic $r-s$ dependence of the term $\cR_{\mu\nu}$. This is telling us that we should replace \Eq{Fmu} with a condition valid only on the slice $r=\b r_*$:
\be\label{Fmu2}
\boxd{\chi_{\mu\nu}(\b r_*,x)=X_{\mu\nu}(\b r_*,x)=R_{\mu\nu}\qquad\Rightarrow\qquad \Phi_{\mu\nu}(\b r_*,x)=G_{\mu\nu}\,.}
\ee
This is the analogue of \Eq{rstcon2}, valid only at $r=\b r_*$. Just like in the scalar system, the auxiliary field $\chi_{\mu\nu}$ is related to the field(s) of the non-local theory by a relation (linear in the Ricci tensor, non-linear in the metric) involving a finite number of derivatives.

On the $r=\b r_*$ slice, the localized and non-local system coincide. Equations \Eq{Xr}, \Eq{slic} and \Eq{Fmu2} imply together \Eq{eomnl2}. All pieces of the puzzle match beautifully. The last one is the equations of motion for the metric, which is \Eq{lasto} (see appendix \ref{app7}). It is immediate to check that this expression agrees with \Eq{nlff3} on the $r=\b r_*$ slice. Using \Eq{difPg}, we can recast all the $s$ integrals in \Eq{lasto} as non-local form factors, $-\int_0^{r_*} \rmd s\,\Phi_{\s\t}(r-s)=\g_{r_*}(\B)\Phi_{\s\t}(r)$. Then, \Eq{slic} does the rest of the job. Also, from \Eq{chimnde} and \Eq{Fmu2}, the last term in \Eq{lasto} reads
\ba
\int_0^{r_*}\!\rmd s\int_0^{s}\!\rmd q\,\bar\Theta_{\mu\nu}[\rme^{-q\B}\chi_{\s\t}(r),\rme^{(q-s)\B}\Phi^{\s\t}(r)] &\stackrel{r=\b r_*}{=}& \int_0^{r_*}\!\rmd s\int_0^{s}\!\rmd q\,\bar\Theta_{\mu\nu}\left[\rme^{-q\B}X_{\s\t},\rme^{(q-s)\B}\Phi^{\s\t}\right]\nonumber\\
&\stackrel{\textrm{\tiny \Eq{app1d2}}}{=}& \Theta_{\mu\nu}(X_{\s\t},\phi^{\s\t}) =\Theta_{\mu\nu}(R_{\s\t},\phi^{\s\t})\,,\nonumber
\ea
thus recovering the last and most complicated piece of \Eq{eomnl1}. Considering the massive effort it takes to derive the non-local Einstein equations \Eq{EinEq1} and \Eq{eomnl1}, the advantage of the diffusion-equation method to write down the dynamics is evident. Here we do not have to deal with the variation of form factors and with the non-commutation rules involving the $\Theta_{\mu\nu}$ and $\vartheta_{\mu\nu}$ functions. Despite being as complicated as the second-order relation \Eq{Theta}, the function $\bar\Theta_{\mu\nu}$ is not nearly as messy as its non-local counterpart \Eq{Theta2app}.


\subsection{Initial conditions and degrees of freedom}\label{lodof}

We are finally in the position to discuss the Cauchy problem of non-local gravity. The localized equations of motion of the previous sub-section are second order in spacetime derivatives for all fields, so that there are only six initial conditions to be specified (``initial'' in spacetime time): $g_{\mu\nu},\dot g_{\mu\nu},\Phi_{\mu\nu},\dot\Phi_{\mu\nu},\chi_{\mu\nu},\dot\chi_{\mu\nu}$. Using the $r=\b r_*$ conditions \Eq{Fmu2}, one can see that $\phi_{\mu\nu}\sim G_{\mu\nu}\sim \p^2 g_{\mu\nu}$ and $\chi_{\mu\nu}\sim R_{\mu\nu}\sim \p^2 g_{\mu\nu}$ in the non-local system, so that, overall, one has to specify only four derivatives of the metric. 

It is well known that the perturbative degrees of freedom are finite, as one can see from the poles of the graviton propagator \cite{Mod1,BGKM,Edh18}. Here we can go beyond that result and make a fully non-perturbative counting. For the graviton there are $D(D+1)/2-D-D=D(D-3)/2$ polarization modes in $D$ dimensions (a symmetric rank-2 tensor with $D(D+1)/2$ components that are not independent due to $D$ Bianchi identities and $D$ diffeomorphisms), both in the localized and in the non-local system. Each of the localized rank-2 symmetric tensor fields $\Phi_{\mu\nu}$ and $\chi_{\mu\nu}$ has $D(D+1)/2$ degrees of freedom, so that the total number of degrees of freedom in the localized system is $D(3D-1)/2$, i.e., 22 in $D=4$. This number is not important, however, because it is greatly reduced in the slice $r=\b r_*$. Therein, $\phi_{\mu\nu}=G_{\mu\nu}$ has $D(D+1)/2-D=D(D-1)/2$ degrees of freedom; the $-D$ comes from the fact that, on shell, the Bianchi identities imply the transverse condition $\N^\mu\Phi_{\mu\nu}=0$. Also, $\phi_{\mu\nu}$ should be regarded as independent of the metric, for the reason that it satisfies non-trivial dynamical equations. The fate of $\chi_{\mu\nu}=R_{\mu\nu}$ is similar to the auxiliary field $\chi$ of the scalar system and consists in getting out of the game on the slice where the localized system reproduces the non-local dynamics. What happens is that $\chi_{\mu\nu}$ depends on the other fields, since from the definition \Eq{Eiten} of the Einstein tensor $\chi_{\mu\nu}=\phi_{\mu\nu}-g_{\mu\nu}\phi/(D-2)$. Therefore, the total number of degrees of freedom of the non-local gravitational system is $D(D-3)/2+D(D-1)/2=D(D-2)$.

The central results of this paper can be summarized as follows.
\be\label{dof3}
\parbox[c]{13cm}{{\bf Number of degrees of freedom: gravity.} \emph{The localized gravitational theory \Eq{gactcm}--\Eq{difeqgcm} in $D+1$ dimensions has $D(3D-1)/2$ degrees of freedom. As a consequence, the non-local gravitational theory \Eq{nlffg} has \underline{$D(D-2)$} non-perturbative degrees of freedom, amounting to \underline{eight} in $D=4$ dimensions.}}
\ee
\be\label{dof2}
\parbox[c]{13cm}{{\bf Number of initial conditions: gravity.} \emph{The Cauchy problem on spacetime slices of the localized gravitational theory \Eq{gactcm}--\Eq{difeqgcm} in $D+1$ dimensions is specified by six initial conditions $g_{\mu\nu}(t_{\rm i},{\bf x})$, $\dot g_{\mu\nu}(t_{\rm i},{\bf x})$, $\Phi_{\mu\nu}(r,t_{\rm i},{\bf x})$, $\dot\Phi_{\mu\nu}(r,t_{\rm i},{\bf x})$, $\chi_{\mu\nu}(r,t_{\rm i},{\bf x})$, $\dot\chi_{\mu\nu}(r,t_{\rm i},{\bf x})$. As a consequence, the non-local non-perturbative gravitational theory \Eq{nlffg} is specified by \underline{four} initial conditions $g_{\mu\nu}(t_{\rm i},{\bf x})$, $\dot g_{\mu\nu}(t_{\rm i},{\bf x})$, $\ddot g_{\mu\nu}(t_{\rm i},{\bf x})$, $\dddot g_{\mu\nu}(t_{\rm i},{\bf x})$.}}
\ee

We have not checked whether $\phi_{\mu\nu}$ can be further decomposed into a spin-2 massive ghost particle with $D(D-1)/2-1$ degrees of freedom and a scalar field, as done in higher-order Stelle gravity \cite{Ste77,Ste78}. Ghost modes are absent at the perturbative level, as proven explicitly in \cite{Mod1,MoRa1,MoRa2}, but their presence at the non-perturbative level remains an open question. We will leave this interesting problem, together with the existence or avoidance of ghosts in the non-local theory \Eq{nlffg}, for the future.

At any rate, we can state that the localized theory does not generate any \emph{extra} non-local ghost problem. It is useful to compare first the non-local theory \Eq{nlffg} with another non-local model employed in cosmology, where the Einstein--Hilbert Lagrangian is modified by a term $R\to \cL\propto R[1+f(\B^{-1}R)]$ \cite{DeWo1}. Auxiliary fields can be introduced so that the Lagrangian $\tilde\cL\propto R[1+f(\Phi)]+\Psi(\B\Phi-R)$ replicates on shell the dynamics of the original system \cite{NoOd}. However, this ``localized'' version is not completely equivalent to the former because the equation of motion $\B\Phi-R=0$ of the Lagrange multiplier $\Psi$ is used to obtain $\Phi$ as a non-local function of $R$. The problem is that the solution of this relation is of the form $\Phi=\B^{-1}R+\la$, where $\la$ is a scalar mode obeying the homogeneous equation $\B\la=0$ \cite{Kos08}. This extra mode is responsible for extending the space of solutions to dynamics not admitted by the original non-local system \cite{Kos08}. Also, it makes an otherwise immaterial ghost degree of freedom dynamical: $\Psi\B\Phi\to-\p_\mu\Psi\p^\mu\Phi=-(1/4)\p_\mu(\Psi+\Phi)\p^\mu(\Psi+\Phi)+(1/4)\p_\mu(\Psi-\Phi)\p^\mu(\Psi-\Phi)$. Suitable conditions on $f$, found along the same lines of ghost constraints in $f(R)$ or higher-order theories \cite{KaSSo,KaSo,NuSo2,Chi05,DHT,CdD}, remove this ghost \cite{DeWo2}. Coming back to the theory studied in the present paper, the localized version \Eq{gactcm}--\Eq{difeqgcm} is not a field redefinition of the model \Eq{nlffg}: it is a different system living in a different number of dimensions that coincides with the non-local system only at a particular slice along the $r$ direction. A second important difference which we already had occasion to appreciate in section \ref{eoms} is that our localization does not entail any homogeneous mode. This means that the ghost mode arising from the mixed kinetic term in \Eq{difeqgcm} and originated by a Lagrange multiplier does not propagate on the $r=\b r_*$ slice.


\section*{Acknowledgments}

\noindent G.C.\ is under a Ram\'on y Cajal contract and thanks Terry Tomboulis for useful e-mail discussions. G.C.\ and L.M.\ are supported by the MINECO I+D grants FIS2014-54800-C2-2-P and FIS2017-86497-C2-2-P.

\appendix


\section{The operator \texorpdfstring{$\B^{-1}$}{}}\label{app1}

In this section, we define the formal expression ``$1/\B$'' and show under what condition
\be\label{b1b}
\B\,\frac{1}{\B} = \frac{1}{\B}\,\B = \mathbbm{1}\,.
\ee
This property allows one to write unordered expressions such as
\be\label{effeb}
\frac{f(\B)}{\B}\,,
\ee
which are used throughout the paper.

Let $\cK(x-y)$ be the solution of the Green equation in a curved $D$-dimensional spacetime,
\be\label{bcK}
\B_x\cK(x-y)=\frac{\de^{(D)}(x-y)}{\sqrt{-g}}\,.
\ee
 Treated as an operator on the space of rapidly decreasing  test functions $\vp$, $\cK$ is nothing but the operator $1/\B$. In fact, define the convolution
\be\label{ckvp}
(\cK\vp)(x):=\int_{-\infty}^{+\infty}\rmd^Dy\,\sqrt{-g}\,\cK(x-y)\,\vp(y)\,.
\ee
Then, from \Eq{bcK} one has
\be\nonumber
\B (\cK\vp)(x)=\int_{-\infty}^{+\infty}\rmd^Dy\,\de^{(D)}(x-y)\,\vp(y)=\vp(x)\,,
\ee
corresponding to $\B\,\B^{-1}\vp=\vp$.

\emph{A priori}, it is not obvious that $\B$ and $\B^{-1}$ commute. Indeed, they do. From \Eq{ckvp}, one has
\ba
\cK\B\vp &=& \int_{-\infty}^{+\infty}\rmd^Dy\,\sqrt{-g}\,\cK(x-y)\,\B_y\vp(y)\nonumber\\
&=&\int_{-\infty}^{+\infty}\rmd^Dy\,\sqrt{-g}\,\B_y\cK(x-y)\,\vp(y)+O(\N)\nonumber\\
&=&\int_{-\infty}^{+\infty}\rmd^Dy\,\sqrt{-g}\,\B_x\cK(x-y)\,\vp(y)+O(\N)\nonumber\\
&\stackrel{\textrm{\tiny \Eq{bcK}}}{=}& \vp(x)+O(\N)\,.
\ea
where $O(\N)$ are boundary terms. Therefore, \Eq{b1b} holds only if boundary terms vanish, which happens if, and only if, $\vp(\pm\infty)=0=\p_\mu\vp(\pm\infty)$. These conditions are always satisfied for rapidly decreasing test functions. In the context of this paper, $\vp$ is a curvature invariant ($R_{\mu\nu}$ or $R$) and the boundary conditions simply require that the curvature and its first derivative vanish at infinity.

An even simpler way to show \Eq{b1b} is the following. Since $\B$ and $\rme^{-r\B}$ commute, then
\be\nonumber
-\int_0^{r_*} \rmd r \,\rme^{-r\B}\B \vp=-\int_0^{r_*} \rmd r \B\,\rme^{-r\B} \vp
\ee
for any test function $\vp$. This equality becomes
\be\nonumber
(\rme^{-r_*\B}-1)\,\frac{1}{\B}\,\B \vp=(\rme^{-r_*\B}-1)\,\B\,\frac{1}{\B} \vp\,,
\ee
which implies \Eq{b1b}.


\section{Original version of the localized scalar system}\label{app2}

For the reader interested in comparing the present formulation of the diffusion-equation method with the original one, we redo the calculation of section \ref{ized} for the localized system presented in \cite{cuta3}. There is a major difference between the simplified system \Eq{act}--\Eq{locch2} and that considered in \cite{cuta3}. All $r_*$ in \Eq{locPh2} and \Eq{locch2}, except in the integration range of $q$, are replaced by $\g r$, where $\g>0$ is a positive constant (not to be confused with the form factor of the present paper). Furthermore, the upper limit of the $q$-integration is now $r$, so that this becomes a nested integral. These changes make calculations slightly more complicated and with no advantage with respect to the easier case of section \ref{ized}. Concretely, equations \Eq{locPh2} and \Eq{locch2} are replaced by
\ba
\cL_{\Phi}&=&\frac12\Phi(r,x)\B \Phi(r-\gamma r,x)-V[\Phi(r,x)]\,,\label{locPh}\\
\cL_{\chi}&=&\frac12\int_0^r \rmd q\,\chi(r'',x)(\gamma\p_{r'}-\B)\Phi(r',x)\,,\label{locch}
\ea
where $r':= r(1-\gamma)+\gamma q$ and $r'':=r-\gamma q$. Varying with respect to $\chi$ yields
\ba
0 &=& \frac{\de\cS[\Phi,\chi]}{\de\chi(\bar r,\bar x)}\Big|_{\bar x=x}= \frac12\int \rmd r \int_0^r \rmd q\,\de(r''-\bar r)(\gamma\p_{r'}-\B)\Phi(r')\nonumber\\
  &=& \frac12\int \rmd r (\gamma\p_{r'}-\B)\Phi(r')\Big|_{r'=r(2-\gamma)-\bar r},\nonumber
\ea
where, for consistency, in the third line we integrated the inner integral in the nested product and, from the integration of the delta, we obtained $r''=\bar r$, hence $\gamma q=r-\bar r$ and $r'=r(2-\gamma)-\bar r$. Having assumed that $0\leq\bar r\leq r$ implies, from $\gamma q=r-\bar r$, that $0\leq q\leq r/\g$, which is trivially satisfied if $\gamma\leq 1$. Thus, we assume (rather than find, as in the new version of the method) that the integrand rather than the integral is zero: $0=(\gamma\p_r-\B)\Phi(r,x)$. Integrating \Eq{locch} by parts,
\ba
\cL_{\chi}&=&\frac12\int_{r(1-\g)}^r\rmd r'\p_{r'}[\chi(r'')\Phi(r')]-\frac12\int_0^r \rmd s\,\Phi(r')(\gamma\p_{r'}+\B)\chi(r'')\nonumber\\
&=&\frac12[\chi(r-\g r)\Phi(r)-\chi(r)\Phi(r-\g r)]-\frac12\int_0^r \rmd s\,\Phi(r')(\gamma\p_{r'}+\B)\chi(r'')\,,\label{inpa}
\ea
so that
\ba
0=\frac{\de\cS[\Phi,\chi]}{\de\Phi(\bar r)}&=&\frac12[\B\Phi(\bar r-\g \bar r)+\chi(\bar r-\g \bar r)]+\frac{1}{2(1-\g)}\left[\B\Phi\left(\frac{\bar r}{1-\g}\right)-\chi\left(\frac{\bar r}{1-\g}\right)\right]\nonumber\\
&&-V'[\Phi(\bar r)]-\frac{1}{2(1-\g)}\int\rmd r\,(\g\p_{\bar r}+\B)\chi(-\bar r+2r-\g r)\,,\label{inter}
\ea
where from the integration of the delta we obtained $r'=\bar r$, hence $\gamma q=\bar r-r(1-\gamma)$ and $r''=r(2-\gamma)-\bar r$. The first condition implies $(1-1/\g) r\leq q\leq r$, which is trivially satisfied if, again $\g\leq 1$. Thus, for self-consistency we must limit the range of values of $\g$ to $0<\g\leq 1$. Thus, the analogues of equations \Eq{eomP12} and \Eq{eomP22} are
\ba
0&=&\frac12[\B\Phi(r-\g r,x)+\chi(r-\g r,x)]+\frac{1}{2(1-\g)}\left[\B\Phi\left(\frac{r}{1-\g},x\right)-\chi\left(\frac{r}{1-\g},x\right)\right]\nonumber\\
&&-V'[\Phi(r,x)]\,,\label{eomP1}\\
0&=&(\g\p_r-\B)\chi(r,x)\,.\label{eomP2}
\ea
Since $\Phi(r-\g r,x)=\rme^{-r\B}\Phi(r,x)$ and $\chi(r-\g r,x)=\rme^{-r\B}\chi(r,x)$ on shell, now equation \Eq{rstcon1} is mandatory if we want to reproduce the exponentials $\exp(-r_*\B)$. On the contrary, in section \ref{ized} we could choose a different $r$ slice where to impose the matching conditions \Eq{rstcon1}. Thus, one recovers \Eq{tpheom} again. In \cite{cuta3}, only the final result was given and the intermediate steps \Eq{inpa}, \Eq{inter} and \Eq{eomP1} were omitted. 


\section{Comments on the system \Eq{tildeL}}\label{appI}

Thanks to \Eq{utile}, the variation of $I$ with respect to $\vp$ vanishes identically:
\ba
\frac{\delta}{\delta\varphi(\bar r)}\int\rmd r\,I(r) &=& \int\rmd r\,[\psi\left(r-\tfrac r_*\right) f_{,\vp}\left(r+\tfrac r_*\right)\delta\left(r+\tfrac r_*-\bar r\right)\nonumber\\
&&-\psi\left(r+\tfrac r_*\right)f_{,\vp}\left(r-\tfrac r_*\right)\delta\left(r-\tfrac r_*-\bar r\right)]\nonumber\\
&&-\int\rmd r\int_0^{r_*}\rmd q\,f_{,\vp}\left(r+q-\tfrac r_*\right)\delta\left(r+q-\tfrac r_*-\bar r\right)\p_q\psi\left(r-q+\tfrac r_*\right)\nonumber\\
&=&\psi(\bar r-r_*)f_{,\vp}(\bar r)-\psi(\bar r+r_*)f_{,\vp}(\bar r)+f_{,\vp}(\bar r)\int_{\bar r-\tfrac r_*}^{\bar r+\tfrac r_*}\rmd r\,\p_r\psi(2r-\bar r)\nonumber\\
&=& f_{,\vp}(\bar r)[\psi(\bar r-r_*)-\psi(\bar r+r_*)+\psi(\bar r+r_*)-\psi(\bar r-r_*)]= 0\,.\nonumber
\ea
On the other hand, the variation of $I$ with respect to $\psi$ is
\ba
\frac{\delta}{\delta\psi(\bar r)}\int\rmd r\,I(r) &=& \int\rmd r\int_0^{r_*}\rmd q\,\delta\left(r-q+\tfrac r_*-\bar r\right)\p_q f\left(r+q-\tfrac r_*\right)\nonumber\\
&&+\psi(\bar r-r_*)f_{,\psi}(\bar r)-\psi(\bar r+r_*)f_{,\psi}(\bar r)\nonumber\\
&&-\int\rmd r\int_0^{r_*}\rmd q\,f_{,\psi}\left(r+q-\tfrac r_*\right)\delta\left(r+q-\tfrac r_*-\bar r\right)\p_q\psi\left(r-q+\tfrac r_*\right)\nonumber\\
&=& f_{,\psi}(\bar r)[\psi(\bar r-r_*)-\psi(\bar r+r_*)]\nonumber\\
&&+\int_{\bar r-\tfrac r_*}^{\bar r+\tfrac r_*}\rmd r\,[\p_r f(2r-\bar r)+f_{,\psi}(\bar r)\p_r\psi(2r-\bar r)]\nonumber\\
&=& f(\bar r+r_*)-f(\bar r-r_*)\,.\nonumber
\ea
Therefore, the equations of motion stemming from the Lagrangian \Eq{tildeL} are
\ba
&& 0 = \B\vp(r)-V'\left[\Phi\left(r-\tfrac r_*\right)\right]\,,\label{eomH1}\\
&& 0 = -\B\psi(r)-V'\left[\Phi\left(r-\tfrac r_*\right)\right]+\frac{1}{r_*}[f(r+r_*)-f(r-r_*)]\,,\label{eomH2}
\ea
were we used \Eq{Pvfpsi}. Taking the difference of \Eq{eomH1} and \Eq{eomH2},
\be\label{Pf}
0=\B\Phi\left(r-\tfrac r_*\right)-\frac{1}{r_*}[f(r+r_*)-f(r-r_*)]\,.
\ee
Comparing with the non-local equation of motion from \Eq{facbis} in terms of $\tilde\phi$, $0=\B\tilde\phi-\rme^{\tfrac r_*\B}V'(\phi)$, \emph{if} the field $\Phi(r,x)$ obeyed the diffusion equation then we could make the identification $\Phi(\b r_*-r_*/2,x)=\tilde\phi(x)$ and we would reproduce the equations of motion if
\be\label{byhand}
\frac{1}{r_*}[f(\b r_*+r_*)-f(\b r_*-r_*)]=\rme^{\tfrac r_*\B}V'[\Phi(\b r_*)]\,.
\ee
As it stands, this equation should hold only at $r=\b r_*$ for some number $\b$.

Here we see two problems with the formulation \Eq{tildeL}. The first is that, having removed the diffusion equation from the dynamics, we have to impose it by hand in order to transform translations in $r$ into non-local exponential operators. This procedure betrays the spirit of the localization procedure, where diffusion was part of the dynamics. However, a diffusion-equation term in the Lagrangian such as $\cL_{\chi}$ necessarily entails a non-diagonal kinetic term, which in turn hides some properties of the degrees of freedom. In order to spell out the nature of these degrees of freedom, we have to diagonalize the kinetic terms, but in doing so the localization procedure cannot be applied transparently. In this case, the match with non-local dynamics entails the imposition of three \emph{ad hoc} conditions (an $r$-dependent diffusion equation for $\vp$ and $\psi$ and the point-wise condition \Eq{byhand}), in comparison with only one condition (the point-wise expression \Eq{rstcon2}).

The second problem has to do with the condition \Eq{byhand}, which was imposed by hand to match the non-local dynamics. The reconstruction of $f$ from \Eq{byhand} may turn out to be very difficult because this condition is assumed at a given point in the $r$-direction. For consistency, one should not attempt to generalize \Eq{byhand} to hold for any $r$. In fact, suppose to replace $\b r_*$ with $r$ everywhere in \Eq{byhand}. If we assume that $f$ (hence both $\Phi$ and $\psi$) translates when exponential operators are applied, then the right-hand side of \Eq{byhand} should do the same. In particular, \Eq{byhand} could be rewritten as $f(r+r_*/2)-f(r-3r_*/2)=r_*V'[\Phi(r)]$. However, $V$ is a non-linear function of $\Phi$ and $V'$ does not obey the diffusion equation unless the potential is quadratic, $V\propto\Phi^2$. Therefore, the approach using \Eq{tildeL} may be unsuitable in the most general case of non-linear self-interactions, unless one considers more complicated functionals $f$.

\section{Variations of curvature invariants and form factors}\label{app3}

This appendix collects a wealth of useful formul\ae\ that can be used when calculating the Einstein equations in any local or non-local metric theory of gravitation. Some of these relations may already be familiar to the reader, but others are delicate and deserve extra care when implemented. For the sake of the record, we do not omit any important detail. Our convention for multiplicative operators with insertions of derivatives will be such that $\N A B\N C=(\N A)B\N C$; differentiation of multiplicative operators to the right will be indicated explicitly with a bracket, $\N (A\N B)$, while $\N\N A B=[\N(\N A)]B$.

First, we recall the index-flipping formula for the variation of the metric (inverse of a matrix),
\be\label{altobasso}
\de g_{\s\t}=-g_{\s\mu}g_{\t\nu}\de g^{\mu\nu}\,,
\ee
and the following one, stemming from \Eq{altobasso} and valid for any symmetric rank-2 tensors $A^{\a\b}$ and $B^{\mu\nu}$:
\be\label{altobasso2}
\de (g_{\a\mu}g_{\b\nu})A^{\a\b}B^{\mu\nu}=-\de (g^{\a\mu}g^{\b\nu})A_{\a\b}B_{\mu\nu}\,.
\ee
Also
\ba
\delta \sqrt{-g} &=& -\frac12\,g_{\mu\nu}\,\sqrt{-g}\,\delta g^{\mu\nu}\,,\label{desg}\\
\de R_{\mu\nu}	 &=& \N^\a\N_{(\mu}\de g_{\nu)\a}-\frac12\left[\B\de g_{\mu\nu}+g^{\a\b}\N_{(\mu}\N_{\nu)}\de g_{\a\b}\right]\,,\label{dRmn}\\
\delta R         &=& \de g^{\mu\nu}R_{\mu\nu} +g^{\mu\nu}\de R_{\mu\nu}\stackbin[\textrm{\tiny \Eq{altobasso}}]{\textrm{\tiny \Eq{dRmn}}}{=} (R_{\mu\nu}+g_{\mu\nu}\,\B -\N_\mu\N_\nu)\,\delta g^{\mu\nu}\,,\label{dRg}
\ea
where $A_{(\mu}B_{\nu)}:=(A_\mu B_\nu+A_\nu B_\mu)/2$. The variation $\de G_{\mu\nu}$ of the Einstein tensor
\be\label{Eiten}
G_{\mu\nu}:= R_{\mu\nu}-\frac12g_{\mu\nu}R
\ee
can be obtained by combining \Eq{dRmn} and \Eq{dRg}. Another expression we will use often is the contraction of $\de R_{\mu\nu}$ with a symmetric rank-2 tensor $A^{\mu\nu}$:
\ba
2\de R_{\mu\nu} A^{\mu\nu} &\stackrel{\textrm{\tiny \Eq{dRmn}}}{=}& 2A^{\mu\nu}\N^\a\N_\mu\de g_{\nu\a}-A^{\mu\nu}\left(\B\de g_{\mu\nu}+g^{\a\b}\N_\mu\N_\nu\de g_{\a\b}\right)\nonumber\\
&=& 2\de g_{\nu\a}\N_\mu\N^\a A^{\mu\nu}-\de g_{\mu\nu}\B A^{\mu\nu}-g^{\a\b}\de g_{\a\b}\N_\mu\N_\nu A^{\mu\nu}+O(\N)\nonumber\\
&\stackrel{\textrm{\tiny \Eq{altobasso}}}{=}& \de g^{\mu\nu}\left(\B A_{\mu\nu}+g_{\mu\nu}\N_\s\N_\t A^{\s\t}-2\N_\s\N_\mu A_\nu^{\ \s}\right)+O(\N)\,,\label{giusta}
\ea
where $O(\N)$ symbolizes total derivative terms that do not contribute to the equations of motions.

Now we list a series of relations involving the variation of the Laplace--Beltrami operator $\B$ and of an arbitrary form factor $\g$. There is one essential fact that one must bear in mind in order to calculate the equations of motion correctly: the metric $g^{\mu\nu}$ does not pass through $\de\B$. The reason is that the operator $\de\B$ acting on a rank-2 tensor of the ``scalar times metric'' form $A g^{\a\b}$ is not the same operator $\de\B$ acting on a scalar, $\de\B (g^{\a\b}A)\neq g^{\a\b}\de\B A$:
\be\label{noncom}
[\de\B,g^{\a\b}]\neq 0\,.
\ee
We calculate the variation of the Laplace--Beltrami operator when acting on an arbitrary rank-2 contravariant symmetric tensor $B^{\a\t}$:
\be\label{deB}
\de\B B^{\a\b}=\de g^{\mu\nu}\N_\mu\N_\nu B^{\a\b}+\de\N_\mu\N^\mu B^{\a\b}+\N^\mu\de\N_\mu B^{\a\b}\,,
\ee
where
\be\label{deB2}
\N_\mu B^{\a\b}=\p_\mu B^{\a\b} +\G^\a_{\mu\rho} B^{\rho\b}+\G^\b_{\mu\rho} B^{\a\rho}\,.
\ee
From \Eq{deB2}, $\de\N_\mu B^{\a\b}=\de\G^\a_{\mu\rho} B^{\rho\b}+\de\G^\b_{\mu\rho} B^{\a\rho}$, where the variation of the Christoffel symbol is $\de\Gamma^\rho_{\a\b}=\frac12 g^{\rho\mu}(\N_\b\de g_{\mu\a}+\N_\a\de g_{\mu\b}-\N_\mu\de g_{\a\b})$. After a tedious calculation, one finds
\ba
\hspace{-.5cm}(\de\B) B^{\a\b} &=& \de g^{\mu\nu}\N_\mu\N_\nu B^{\a\b}+\N_\nu\de g^{\mu\nu}\N_\mu B^{\a\b}-\frac12 g_{\mu\nu} \N^\rho\de g^{\mu\nu}\N_\rho B^{\a\b}\nonumber\\
									 && -g_{\mu}^{(\a} B^{\b)}_{\ \nu}\B\de g^{\mu\nu}-g_{\nu}^{(\a} B^{\b)\rho}\N_\mu\N_\rho\de g^{\mu\nu}+B_{\mu}^{(\b}\N_\nu\N^{\a)}\de g^{\mu\nu}\nonumber\\
									 && +2\left[\N_\nu B^{(\a}_\mu \N^{\b)}\de g^{\mu\nu}-g_\nu^{(\a}\N_\rho B^{\b)}_{\ \mu}\N^\rho\de g^{\mu\nu}-\N_\mu B^{\rho(\a} g^{\b)}_{\nu}\N_\rho\de g^{\mu\nu}\right].\label{c5new}
\ea
Because of \Eq{noncom}, $A_{\a\b}\de\B B^{\a\b}\neq A^{\a\b}\de\B B_{\a\b}$. We also report the variation
\ba
\hspace{-.5cm}(\de\B) B_{\a\b} &=& \de g^{\mu\nu}\N_\mu\N_\nu B_{\a\b}+\N_\nu\de g^{\mu\nu}\N_\mu B_{\a\b}-\frac12 g_{\mu\nu} \N^\rho\de g^{\mu\nu}\N_\rho B_{\a\b}\nonumber\\
									 && +g_{\mu(\a} B_{\b)\nu}\B\de g^{\mu\nu}-g_{\nu(\a} B_{\b)\rho}\N_\mu\N^\rho\de g^{\mu\nu}+B_{\mu(\b}\N_\nu\N_{(\a}\de g^{\mu\nu}\nonumber\\
									 && +2\left[\N_\nu B_{\mu(\a} \N_{\b)}\de g^{\mu\nu}+g_{\nu(\a}\N_\rho B_{\b)\mu}\N^\rho\de g^{\mu\nu}-\N_\mu B_{\rho(\a} g_{\b)\nu}\N^\rho\de g^{\mu\nu}\right].\label{c5}
\ea
For a generic symmetric rank-2 tensor $A_{\a\b}$, integrating by parts \Eq{c5new} gives
\ba
A_{\a\b}(\de\B) B^{\a\b} &=& \de g^{\mu\nu}A_{\a\b}\N_\mu\N_\nu B^{\a\b}+\N_\nu\de g^{\mu\nu}A_{\a\b}\N_\mu B^{\a\b}-\frac12 g_{\mu\nu} \N^\rho\de g^{\mu\nu}A_{\a\b}\N_\rho B^{\a\b}\nonumber\\
									 &&\stackrel{\downarrow}{-}A_{\mu\b}\left(B^\b_{\ \nu}\B\de g^{\mu\nu}+B^{\b\rho}\N_\nu\N_\rho\de g^{\mu\nu}\right)+A_{\a\b}B^\b_{\ \mu}\N_\nu\N^\a\de g^{\mu\nu}\nonumber\\
									 && +2\left(A_{\a\b}\N_\nu B^\a_{\ \mu} \N^\b\de g^{\mu\nu}\stackrel{\downarrow}{-}A_{\b\nu}\N_\rho B^\b_{\ \mu}\N^\rho\de g^{\mu\nu}-A_{\nu\a}\N_\mu B^{\rho\a}\N_\rho\de g^{\mu\nu}\right)\nonumber\\
									 &=& \de g^{\mu\nu}\big[A_{\a\b}\N_\mu\N_\nu B^{\a\b}-\N_\nu(A_{\a\b}\N_\mu B^{\a\b})+\frac12 g_{\mu\nu} \N_\rho(A_{\a\b}\N^\rho B^{\a\b})\nonumber\\
									 && \stackrel{\downarrow}{-}\B(A_{\mu\b}B^\b_{\ \nu})-\N_\a\N_\mu(A_{\nu\b}B^{\b\a})+\N^\b\N_\nu(A_{\a\b}B^\a_{\ \mu})\nonumber\\
									 && -2\N^{\b}(A_{\a\b}\N_\nu B^\a_{\ \mu})\stackrel{\downarrow}{+}2\N_\a(A_{\nu\b}\N^\a B^\b_{\ \mu})+2\N_\a(A_{\nu\b}\N_\mu B^{\a\b})\big]+O(\N)\nonumber\\
									 &=&\de g^{\mu\nu}\left[-\N_\mu A_{\a\b}\N_\nu B^{\a\b}+\frac12 g_{\mu\nu} \N_\rho(A_{\a\b}\N^\rho B^{\a\b})\right.\no 
&&\stackrel{\downarrow}{+}\N_\a(A_{\mu\b}\N^\a B^\b_{\ \nu}-B^\b_{\ \nu}\N^\a A_{\mu\b})+\N_{\b}(B_{\mu\a}\N_\nu A^{\a\b}-A^{\a\b}\N_\nu B_{\mu\a})\no
&&\left.\vphantom{\frac12}+\N_\a(A_{\mu\b}\N_\nu B^{\b\a}-B^{\b\a}\N_\nu A_{\mu\b})\right]+O(\N)\nonumber\\
									&=& \de g^{\mu\nu}\bar\Theta_{\mu\nu}(A_{\a\b},B^{\a\b})+O(\N)\,,\label{dbmn}
\ea
where
\ba
\bar\Theta_{\mu\nu}(A_{\a\b},B^{\a\b}) &:=& \bar\Theta_{\mu\nu}^{\rm sym}(A_{\a\b},B^{\a\b})+\bar\Theta_{\mu\nu}^{\rm antisym}(A_{\a\b},B^{\a\b})\,,\label{Theta}\\
\bar\Theta_{\mu\nu}^{\rm sym}(A_{\a\b},B^{\a\b})&:=& -\N_\mu A_{\a\b}\N_\nu B^{\a\b}+\frac14 g_{\mu\nu} \N_\rho(A_{\a\b}\N^\rho B^{\a\b}+B^{\a\b}\N^\rho A_{\a\b})\,,\\
\bar\Theta_{\mu\nu}^{\rm antisym}(A_{\a\b},B^{\a\b})&:=&\frac14 g_{\mu\nu} \N_\rho(A_{\a\b}\N^\rho B^{\a\b}-B^{\a\b}\N^\rho A_{\a\b})\stackrel{\downarrow}{+}\N_\a(A_{\mu\b}\N^\a B^\b_{\ \nu}-B^\b_{\ \nu}\N^\a A_{\mu\b})\nonumber\\
									 && +\N_{\b}(B_{\mu\a}\N_\nu A^{\a\b}-A^{\a\b}\N_\nu B_{\mu\a})+\N_\a(A_{\mu\b}\N_\nu B^{\b\a}-B^{\b\a}\N_\nu A_{\mu\b})\,.\no			
\ea
It is not difficult to prove similar expressions from \Eq{c5}, identical except for the signs marked with an arrow, which are flipped. In particular, the function $\bar\Theta_{\mu\nu}$ is not symmetric under index lowering and raising of its arguments:
\be\label{TT}
\bar\Theta_{\mu\nu}(A_{\a\b},B^{\a\b})=\bar\Theta_{\mu\nu}(A^{\a\b},B_{\a\b})+2\N_\a(A_{\mu\b}\N^\a B^\b_{\ \nu}-B^\b_{\ \nu}\N^\a A_{\mu\b})\,.
\ee
For two scalars, it is easy to see that
\ba
A (\de\B) B &=& \de g^{\mu\nu}\left[-\N_\mu B\N_\nu A+\frac12 g_{\mu\nu} \N_\rho(A\N^\rho B)\right]+O(\N)\nonumber\\
						&=&\de g^{\mu\nu}\bar\vartheta_{\mu\nu}(A,B)+O(\N)\,,\label{scs}
\ea
where
\be\label{thet}
\bar\vartheta_{\mu\nu}(A,B):=-\N_\mu  B\N_\nu A+\frac12 g_{\mu\nu} (A\B B+\N_\rho A\N^\rho B)\,.
\ee
When $A_{\a\b}=A g_{\a\b}$ for some scalar $A$, the following important formula (up to total derivatives) can be derived from \Eq{Theta}:
\be\label{5star0}
A g_{\a\b}\de\B B^{\a\b}-A\de\B (g_{\a\b} B^{\a\b})=\de g^{\mu\nu}(A\B B_{\mu\nu}-B_{\mu\nu}\B A)\,.
\ee
This equation is the proof of \Eq{noncom}. Also, replacing $A_{\a\b}=A g_{\a\b}$ or $B^{\a\b}=B g^{\a\b}$ into \Eq{Theta}, we get two independent formul\ae\
\ba
\bar\Theta_{\mu\nu}(A g_{\a\b},B^{\a\b}) &=& \bar\vartheta_{\mu\nu}(A,B)+A\B B_{\mu\nu}-B_{\mu\nu}\B A\,,\qquad B=g_{\a\b}B^{\a\b}\,,\label{usef1}\\
\bar\Theta_{\mu\nu}(A_{\a\b},B g^{\a\b}) &=& \bar\vartheta_{\mu\nu}(A,B)+A_{\mu\nu}\B B-B\B A_{\mu\nu}\,,\qquad A=g^{\a\b}A_{\a\b}\,,\label{usef2}
\ea
which cannot be combined together unless $B^{\a\b}=A^{\a\b}$. 
 Similar expressions are obtained when lowering and raising the indices $\a\b$, but with opposite sign of the second and third terms of \Eq{usef1} and \Eq{usef2}.

We generalize the above formul\ae\ for an arbitrary form factor \Eq{ggen}. Since $\de\g=\sum_{n=1}^{+\infty}\sum_{k=1}^n c_n\B^{k-1}\de\B\B^{n-k}$, it is not difficult to find
\be\label{usef4}
A_{\a\b}\de\g B^{\a\b}=\de g^{\mu\nu}\Theta_{\mu\nu}(A_{\a\b},B^{\a\b})\,,\qquad A\de\g B=\de g^{\mu\nu}\vartheta_{\mu\nu}(A,B)\,,
\ee
up to total derivatives, where
\ba
\Theta_{\mu\nu}(A_{\a\b},B^{\a\b})&=&\sum_{n=1}^{+\infty}\sum_{k=1}^n c_n\bar\Theta_{\mu\nu}(\B^{k-1}A_{\a\b},\B^{n-k}B^{\a\b})\,,\label{usef5a}\\
\vartheta_{\mu\nu}(A,B)&=&\sum_{n=1}^{+\infty}\sum_{k=1}^n c_n\bar\vartheta_{\mu\nu}(\B^{k-1}A,\B^{n-k}B)\,.\label{usef5b}
\ea
Both pairs $(\bar\Theta_{\mu\nu},\bar\vartheta_{\mu\nu})$ and $(\Theta_{\mu\nu},\vartheta_{\mu\nu})$ are bilinear functionals of their arguments. By virtue of \Eq{TT}, 
\be\label{TT2}
A_{\a\b}\de\g B^{\a\b}\neq A^{\a\b}\de\g B_{\a\b}\,,
\ee
as is obvious when considering the terms of the variation $$\de(g^{\mu\a}g^{\nu\b}A_{\a\b}\g B_{\mu\nu})=\de[A_{\a\b}\g (g^{\mu\a}g^{\nu\b}B_{\mu\nu})].$$ Also, from \Eq{usef4} we get
\ba
A g_{\a\b}\de\g B^{\a\b} &\stackrel{\textrm{\tiny \Eq{usef5a}}}{=}& \de g^{\mu\nu}\sum_{n=1}^{+\infty}\sum_{k=1}^n c_n\bar\Theta_{\mu\nu}(g_{\a\b}\B^{k-1}A,\B^{n-k}B^{\a\b})+O(\N)\nonumber\\
&\stackrel{\textrm{\tiny \Eq{usef1}}}{=}& A\de\g B+\de g^{\mu\nu}\sum_{n=1}^{+\infty}\sum_{k=1}^n c_n(\B^{k-1}A\B^{n-k+1}B_{\mu\nu}-\B^{n-k}B_{\mu\nu}\B^k A)+O(\N)\nonumber\\
&=& A\de\g B\nonumber\\
&&+\de g^{\mu\nu}\sum_{n=1}^{+\infty} c_n\left(A\B^n B_{\mu\nu}-\B^{n-1}B_{\mu\nu}\B A+\B^{n-1}A\B B_{\mu\nu}-B_{\mu\nu}\B^n A\right)\nonumber\\
&& +\de g^{\mu\nu}\sum_{n=1}^{+\infty} c_n\sum_{k=2}^{n-1}(\B^{k-1}A\B^{n-k+1}B_{\mu\nu}-\B^{n-k}B_{\mu\nu}\B^k A)+O(\N)\nonumber\\
&=& A\de\g B+\de g^{\mu\nu}(A\g B_{\mu\nu}-B_{\mu\nu}\g A)\nonumber\\
&&+\de g^{\mu\nu}\sum_{n=1}^{+\infty} c_n\left(\B^{n-1}A\B B_{\mu\nu}-\B^{n-1}B_{\mu\nu}\B A\right)\nonumber\\
&& +\de g^{\mu\nu}\sum_{n=1}^{+\infty} c_n\sum_{k=2}^{n-1}(\B^{k-1}A\B^{n-k+1}B_{\mu\nu}-\B^k A\B^{n-k}B_{\mu\nu})+O(\N)\nonumber\\
&=& A\de\g B+\de g^{\mu\nu}(A\g B_{\mu\nu}-B_{\mu\nu}\g A)+\de g^{\mu\nu}\sum_{n=1}^{+\infty} c_n\sum_{k=2}^{n}\B^{k-1}A\B^{n-k+1}B_{\mu\nu}\nonumber\\
&& -\de g^{\mu\nu}\sum_{n=1}^{+\infty} c_n\sum_{k=1}^{n-1}\B^k A\B^{n-k}B_{\mu\nu}+O(\N)\nonumber\\
&\stackrel{k'=k-1}{=}& A\de\g B+\de g^{\mu\nu}(A\g B_{\mu\nu}-B_{\mu\nu}\g A)+\de g^{\mu\nu}\sum_{n=1}^{+\infty} c_n\sum_{k'=1}^{n-1}\B^{k'}A\B^{n-k'}B_{\mu\nu}\nonumber\\
&& -\de g^{\mu\nu}\sum_{n=1}^{+\infty} c_n\sum_{k=1}^{n-1}\B^k A\B^{n-k}B_{\mu\nu}+O(\N)\nonumber\\
&=&A\de\g B+\de g^{\mu\nu}(A\g B_{\mu\nu}-B_{\mu\nu} \g A)+O(\N)\,.\label{5star}
\ea
Analogously, $A_{\a\b}\de\g (B g^{\a\b})=A\de\g B+\de g^{\mu\nu}(A_{\mu\nu} \g B-B\g A_{\mu\nu})+O(\N)$. Therefore, for any form factor
\ba
\Theta_{\mu\nu}(A g_{\a\b},B^{\a\b})&=&\vartheta_{\mu\nu}(A,B)+A\g B_{\mu\nu}-B_{\mu\nu} \g A\,,\label{usef7a}\\
\Theta_{\mu\nu}(A_{\a\b},B g^{\a\b})&=&\vartheta_{\mu\nu}(A,B)+A_{\mu\nu}\g B-B\g A_{\mu\nu}\,.\label{usef7b}
\ea
In particular,
\be\label{RGGR}
\Theta_{\mu\nu}(G_{\a\b},R^{\a\b})-\Theta_{\mu\nu}(R_{\a\b},G^{\a\b})=R_{\mu\nu}\g R-R\g R_{\mu\nu}\,.
\ee
Moreover, if $B^{\a\b}=A g^{\a\b}$, from \Eq{5star} we obtain
\be\label{usef6}
(A g_{\a\b})\de\g(A g^{\a\b})=D\,\!A\de\g A+O(\N)\qquad\Rightarrow\qquad \Theta_{\mu\nu}(A g_{\a\b},A g^{\a\b})=D\vartheta_{\mu\nu}(A,A)\,.
\ee


\section{Derivation of the Einstein equations \Eq{EinEq1}}\label{app4}

The Einstein equations \Eq{EinEq1} can be obtained through the following steps. First, we note that
\ba
\de(G_{\mu \nu} \g R^{\mu\nu})&=& \delta \left(G_{\mu \nu} \g G^{\mu\nu}-\frac{1}{D-2} G \g G\right)\nonumber\\
&=&\delta \left(g^{\alpha \mu} g^{\beta \nu} \right)  G_{\alpha \beta} \g  G_{\mu\nu}+2\delta G_{\mu \nu} \g G^{\mu\nu} - \frac{2}{D-2}\delta{G} \, \g G + G_{\mu \nu} \, \delta \g G^{\mu\nu} \nonumber\\
&&-\frac{1}{D-2} G  \delta \g G+O(\N)\nonumber\\
&=&\delta \left(g^{\alpha \mu} g^{\beta \nu} \right)  G_{\alpha \beta} \, \g  G_{\mu\nu}+ 2 \delta G_{\mu \nu} \, \g \, G^{\mu\nu}+\delta R \, \g G+G_{\mu \nu} \, \delta \g G^{\mu\nu}\no
&&-\frac{1}{D-2} G  \delta \g G+O(\N)\nonumber\\
&\stackrel{\textrm{\tiny \Eq{altobasso}}}{=}& \delta \left(g^{\alpha \mu} g^{\beta \nu} \right)  G_{\alpha \beta} \, \g  G_{\mu\nu} +2\de R_{\mu\nu}\g G^{\mu\nu}+\de g^{\mu\nu}R\g G_{\mu\nu}\nonumber\\
&& +G_{\mu \nu} \, \delta \g G^{\mu\nu}-\frac{1}{D-2} G  \delta \g G+O(\N),
\label{giusta3}
\ea
where $G$ is the trace of the Einstein tensor. From \Eq{usef4}, the last line is proportional to $\de g^{\mu\nu}$ with coefficient
\ba
\Theta_{\mu\nu}(G_{\s\t},G^{\s\t})-\frac{1}{D-2}\vartheta_{\mu\nu}(G,G) &=& \Theta_{\mu\nu}(G_{\s\t},G^{\s\t})+\frac{1}{2}\vartheta_{\mu\nu}(R,G)\nonumber\\
&=& \Theta_{\mu\nu}(R_{\s\t},G^{\s\t})+\frac12\left[\vartheta_{\mu\nu}(R,G)-\Theta_{\mu\nu}(g_{\s\t}R,G^{\s\t})\right]\nonumber\\
&\stackrel{\textrm{\tiny \Eq{usef7a}}}{=}& \Theta_{\mu\nu}(R_{\s\t},G^{\s\t})+\frac12(G_{\mu\nu} \g R-R\g G_{\mu\nu}).\label{giusta4}
\ea
Then,
\ba
\hspace{-.8cm}\frac{2\kappa^2}{\sqrt{-g}}\de (\sqrt{-g}\,\cL_g) &=& \frac{\delta\sqrt{-g}}{\sqrt{-g}} \left(R-2\Lambda+ G_{\mu \nu} \g R^{\mu\nu} \right) +\delta R+\delta\left(G_{\mu\nu} \g R^{\mu\nu} \right) \nonumber\\
&\stackrel{\textrm{\tiny \Eq{giusta3}}}{=}&\delta g^{\mu\nu} \Big(\underbrace{\vphantom{\frac2D}G_{\mu\nu}+\Lambda g_{\mu\nu}}_{\squared{1}} \underbrace{-\frac{1}{2}g_{\mu\nu}\,G_{\alpha \beta} \, \g R^{\alpha \beta}}_{\squared{2}} \Big)\underbrace{+\delta \left(g^{\alpha \mu} g^{\beta \nu} \right)  G_{\alpha \beta} \, \g  G_{\mu\nu} \vphantom{\frac12}}_{\squared{3}} \nonumber\\
&& \underbrace{+\vphantom{\frac2D} 2\de R_{\mu\nu}\g G^{\mu\nu}}_{\squared{4}}\underbrace{+\vphantom{\frac2D}\frac12\de g^{\mu\nu}(G_{\mu\nu} \g R+R\g G_{\mu\nu})}_{\squared{5}}\underbrace{+\vphantom{\frac2D}R_{\mu \nu} \, \delta \g G^{\mu\nu}}_{\squared{6}}+O(\N).\label{variationAap}
\ea
We labeled the terms with numbers from $\squared{1}$ to $\squared{6}$ for later use; this expression will be extremely useful to prove the equivalence with the non-local system with auxiliary field $\phi^{\mu\nu}$. Using \Eq{giusta},
\ba
\frac{2\kappa^2}{\sqrt{-g}}\de (\sqrt{-g}\,\cL_g) &=& \delta g^{\mu\nu} \left(G_{\mu\nu}+\Lambda g_{\mu\nu} - \frac{1}{2}  g_{\mu\nu} \, G_{\alpha \beta} \g R^{\alpha \beta} \right)+2\de g^{\mu\nu}G^\s_{\ \mu} \g  G_{\nu\s}\nonumber\\
&& +\de g^{\mu\nu}\left(\g\B G_{\mu\nu}+g_{\mu\nu}\N^\s\N^\t\g G_{\s\t}-2\N^\s\N_{\mu}\g G_{\nu\s}\right)\nonumber\\
&&+\frac12\de g^{\mu\nu}(G_{\mu\nu} \g R+R\g G_{\mu\nu})+\de g^{\mu\nu}\Theta_{\mu\nu}(R_{\s\t},G^{\s\t})+O(\N)\,,
\ea
which yields \Eq{EinEq1}. One should be careful about the order of derivative operators acting on Riemann invariants. For instance, $[\N_\mu,\B]R=R_{\mu\nu}\N^\nu R$. To avoid getting entangled with commutation formul\ae, we have kept the natural order of the operators as it came from the variation of the action.

When the form factor $\g$ is chosen to be \Eq{fofag}, the functions $\Theta_{\mu\nu}$ and $\vartheta_{\mu\nu}$ can be written as parametric integrals rather than sums. The variation of the form factor is based upon Duhamel's identity \cite{Yan02}
\be\label{para}
\de \rme^{-s\B} = -\int_0^{s} \rmd q\, \rme^{-q\B}(\de\B)\rme^{(q-s)\B}\,,\qquad q\leftrightarrow s-q\,,
\ee
which is invariant under a reparametrization $q\to q'=s-q$. 
Noting that \Eq{fofag} can be expressed as
\be\label{fofag2}
\g_{r_*}(\B) =  \frac{\rme^{-r_*\B}-1}{\Box}=-\int_0^{r_*}\rmd s\,\rme^{-s\B}\,,
\ee
one can show explicitly that the variation operation commutes with the integral, $\de\g_{r_*}=-\int_0^{r_*}\rmd s\,\de\rme^{-s\B}$. Then we have
\ba
A_{\s\t}\de\g B^{\s\t} &\stackrel{\textrm{\tiny \Eq{fofag2}}}{=}& -\int_0^{r_*}\rmd s\,A_{\s\t}\,\de\rme^{-s\B}B^{\s\t}\nonumber\\
&\stackrel{\textrm{\tiny \Eq{para}}}{=}& \int_0^{r_*}\rmd s\int_0^{s} \rmd q\, A_{\s\t}\,\rme^{-q\B}(\de\B)\rme^{(q-s)\B}B^{\s\t}\nonumber\\
&=& \int_0^{r_*}\rmd s\int_0^{s} \rmd q\, \rme^{-q\B} A_{\s\t}(\de\B)\rme^{(q-s)\B}B^{\s\t}+O(\N)\nonumber\\
&\stackrel{\textrm{\tiny \Eq{dbmn}}}{=}& \de g^{\mu\nu}\int_0^{r_*}\rmd s\int_0^s \rmd q\,\bar\Theta_{\mu\nu}[\rme^{-q\B} A_{\s\t},\rme^{(q-s)\B}B^{\s\t}]+O(\N)\nonumber\\
&=:&\de g^{\mu\nu}\Theta_{\mu\nu}(A_{\s\t},B^{\s\t})+O(\N)\,,\label{app1d2}\\
A\,(\de\g)\,B &\stackrel{\textrm{\tiny \Eq{fofag2}}}{=}& -\int_0^{r_*}\rmd s\,A\,\de\rme^{-s\B}B\nonumber\\
&\stackrel{\textrm{\tiny \Eq{para}}}{=}& \int_0^{r_*}\rmd s\int_0^{s} \rmd q\, A\,\rme^{-q\B}(\de\B)\rme^{(q-s)\B}B\nonumber\\
&=& \int_0^{r_*}\rmd s\int_0^{s} \rmd q\, \rme^{-q\B}A\,(\de\B)\rme^{(q-s)\B}B+O(\N)\nonumber\\
&\stackrel{\textrm{\tiny \Eq{scs}}}{=}& \de g^{\mu\nu}\int_0^{r_*}\rmd s\int_0^s \rmd q\,\bar\vartheta_{\mu\nu}[\rme^{-q\B}A,\rme^{(q-s)\B}B]+O(\N)\nonumber\\
&=:& \de g^{\mu\nu}\vartheta_{\mu\nu}(A,B)+O(\N)\,.\label{app1e2}
\ea
Noting that
\be\label{intint}
\int_0^{r_*}\rmd s \int_0^s \rmd q\,f(s,q)=\int_0^{r_*}\rmd q \int_q^{r_*} \rmd s\,f(s,q)\,,
\ee
for any $f(s,q)$, we obtain the final expressions
\ba
\Theta_{\mu\nu}(A_{\s\t},B^{\s\t}) &=&-\int_0^{r_*}\rmd q\,\bar\Theta_{\mu\nu}[\rme^{-q\B} A_{\s\t},\g_{r_*-q}(\B)B^{\s\t}]\,,\label{Theta2app}\\
\vartheta_{\mu\nu}(A,B) &=& -\int_0^{r_*}\rmd q\,\bar\vartheta_{\mu\nu}[\rme^{-q\B}A,\g_{r_*-q}(\B)B]\,.
\ea
Both $\Theta_{\mu\nu}$ and $\vartheta_{\mu\nu}$ are symmetric in $\mu\nu$. Also, remember from \Eq{TT} and \Eq{TT2} that $\Theta_{\mu\nu}(A_{\s\t},B^{\s\t})\neq \Theta_{\mu\nu}(A^{\s\t},B_{\s\t})$, while $\vartheta_{\mu\nu}(A,A)$ is obviously symmetric with respect to the exchange of its arguments, since it is unambiguously defined via the invariance $q\leftrightarrow s-q$ of Duhamel's formula \Eq{para}.


\section{Derivation of equation \Eq{eomnl1}}\label{app5}

In this section, we derive the equations of motion \Eq{eomnl1} stemming from \Eq{nlff3} in the presence of matter and for an arbitrary form factor $\g$ with trivial kernel. When varying the action, we decide that the contravariant symmetric tensor $\phi^{\mu\nu}$ is constant in the metric, while $\phi_{\mu\nu}$ must be varied. Of course, this is only a convention to make the calculation consistent, but the final result will be the same. Consequently, using
\ba
\de(\phi\g\phi)&=&\de(g_{\a\b}g_{\mu\nu}\phi^{\a\b}\g\phi^{\mu\nu})=\de g_{\a\b}\phi^{\a\b}\g\phi+\de g_{\mu\nu}\phi\g\phi^{\mu\nu}+g_{\mu\nu}\phi\de\g\phi^{\mu\nu}\nonumber\\
&=&\de g_{\mu\nu}(\phi^{\mu\nu}\g\phi+\phi\g\phi^{\mu\nu})+g_{\mu\nu}\phi\de\g\phi^{\mu\nu}\nonumber\\
&\stackrel{\textrm{\tiny \Eq{altobasso}}}{=}&-\de g^{\mu\nu}(\phi_{\mu\nu}\g\phi+\phi\g\phi_{\mu\nu})+g_{\mu\nu}\phi\de\g\phi^{\mu\nu}\,,\label{giusta2}
\ea
where $\phi=g_{\mu\nu}\phi^{\mu\nu}$ is the trace, the variation $\de \tilde S[g,\phi]/\de g^{\mu\nu}$ is
\ba
\hspace{-.5cm}\frac{1}{\sqrt{-g}}\de (\sqrt{-g}\,\tilde\cL[g,\phi]) &=& \frac{\de \sqrt{-g}}{\sqrt{-g}}\tilde \cL[g,\phi]+\de \tilde\cL[g,\phi]\stackrel{\textrm{\tiny \Eq{desg}}}{=} -\frac12\,g_{\mu\nu}\,\delta g^{\mu\nu} \tilde\cL[g,\phi]+\de \tilde\cL[g,\phi]\nonumber\\
											&\stackbin[\textrm{\tiny \Eq{giusta2}}]{\textrm{\tiny \Eq{dRg}}}{=}& \delta g^{\mu\nu}\Big(G_{\mu\nu}+\Lambda g_{\mu\nu}-\frac12 g_{\mu\nu} X_{\s\t}\g\phi^{\s\t}\Big)-\de (g_{\mu\s}g_{\nu\t})\phi^{\s\t}\g\phi^{\mu\nu}\nonumber\\
											 && +2\de R_{\mu\nu}\g\phi^{\mu\nu}-\frac{1}{D-2}\de g^{\mu\nu}(\phi_{\mu\nu}\g\phi+\phi\g\phi_{\mu\nu})+X_{\mu\nu}\de\g\phi^{\mu\nu}\nonumber\\	
										 	 &\stackrel{\textrm{\tiny \Eq{altobasso2}}}{=}& \delta g^{\mu\nu}\Big(\underbrace{\vphantom{\frac12}G_{\mu\nu}+\Lambda g_{\mu\nu}}_{\circled{1}}\underbrace{-\frac12 g_{\mu\nu} X_{\s\t}\g\phi^{\s\t}}_{\circled{2}}\Big)\underbrace{\vphantom{\frac12}+\de (g^{\mu\s}g^{\nu\t})\phi_{\s\t}\g\phi_{\mu\nu}}_{\circled{3}}\nonumber\\
											 && \underbrace{+2\de R_{\mu\nu}\g\phi^{\mu\nu}\vphantom{\frac1D}}_{\circled{4}}\underbrace{-\frac{1}{D-2}\de g^{\mu\nu}(\phi_{\mu\nu}\g\phi+\phi\g\phi_{\mu\nu})}_{\circled{5}}\underbrace{+X_{\mu\nu}\de\g\phi^{\mu\nu}\vphantom{\frac1D}}_{\circled{6}}\,,\label{varap2}
\ea
where $X_{\mu\nu}=2R_{\mu\nu}-\phi_{\mu\nu}+g_{\mu\nu}\phi/(D-2)$ was defined in \Eq{X}. Applying \Eq{giusta}, \Eq{usef5a} and \Eq{usef5b} and taking into account the energy-momentum tensor \Eq{emt}, one obtains \Eq{eomnl1}.


\section{Equivalence of equations \Eq{EinEq1} and \Eq{eomnl1}}\label{app6}

We can make this check by using the formal variations \Eq{variationAap} and \Eq{varap2}. Plugging \Eq{eomnl2} into \Eq{varap2}, one has
\ba
\circled{1} &=& G_{\mu\nu}+\Lambda g_{\mu\nu} = \squared{1}\,,\nonumber\\
\circled{2} &=& -\frac12 g_{\mu\nu} R_{\s\t}\g G^{\s\t}= -\frac12 g_{\mu\nu}\,G_{\s\t}\g R^{\s\t}= \squared{2}\,,\nonumber\\
\circled{3} &=& \de (g^{\mu\s}g^{\nu\t})G_{\s\t}\g G_{\mu\nu}= \squared{3}\,,\nonumber\\
\circled{4} &=& 2\de R_{\mu\nu}\g G^{\mu\nu} =\squared{4}\,,\nonumber\\
\circled{5} &=& -\frac{1}{D-2}\de g^{\mu\nu}(G_{\mu\nu}\g G+G\g G_{\mu\nu})= \frac{1}{2}\de g^{\mu\nu}(G_{\mu\nu}\g R+R\g G_{\mu\nu})=\squared{5}\,,\nonumber\\
\circled{6} &=& R_{\mu\nu}\de\g G^{\mu\nu}=\squared{6}\,.\nonumber
\ea


\section{Derivation of equation \Eq{lasto}}\label{app7}

We split \Eq{gactcm} into separate contributions, $\cS_g=\cS_R+\cS_\Phi+\cS_\chi+\cS_\la$. Variation of the last term yields
\be
0 = 2\k^2\frac{\de\cS_\la}{\de g^{\mu\nu}(\bar r)}=\frac{1}{\sqrt{g(\bar r)}}\p_{\bar r}[\sqrt{g(\bar r)}\la_{\mu\nu}(\bar r)]-\frac12\,g_{\mu\nu}(\bar r)\cL_\la \stackrel{\textrm{\tiny \Eq{eomla}}}{=} \p_r\la_{\mu\nu}\,,
\ee
which will play no part in the equations of motion if we just choose a gauge where the Lagrange multiplier $\la_{\mu\nu}$ is $r$-independent:
\be
\la_{\mu\nu}(r,x)=\la_{\mu\nu}(x)\qquad \textrm{(gauge fixing)}\,.
\ee
Variation of $\cS_R$ gives the Einstein tensor plus a cosmological constant. Next,
\ba
2\k^2\frac{\de\cS_\chi}{\de g^{\mu\nu}} &\stackbin[\textrm{\tiny \Eq{difPg}}]{\textrm{\tiny \Eq{eomla}}}{=}& \int\rmd^Dx\,\rmd r\,\sqrt{-g}\int_0^{r_*}\rmd s\int_0^{s}\rmd q\,\chi_{\mu\nu}(r-q)\frac{\de\B}{\de g^{\mu\nu}}\Phi^{\mu\nu}(r')\nonumber\\
&\stackrel{\textrm{\tiny \Eq{dbmn}}}{=}& \int_0^{r_*}\rmd s\int_0^{s}\rmd q\,\bar\Theta_{\mu\nu}[\chi_{\s\t}(r-q),\Phi^{\s\t}(r+q-s)]\,.
\ea
Finally,
\ba
2\k^2\frac{\de\cS_\Phi}{\de g^{\mu\nu}} &=& \int\rmd^Dx\,\rmd r\,\sqrt{-g}\int_0^{r_*}\rmd s\left\{-\frac{\de(g^{\s\s'}g^{\t\t'})}{\de g^{\mu\nu}}\left[\vphantom{\frac{1}{D-2}}\Phi_{\s\t}(r)\Phi_{\s'\t'}(r-s)\right.\right.\nonumber\\
&&\left.\left.-\frac{1}{D-2}\Phi^{\s\s'}(r)\Phi^{\t\t'}(r-s)\right]-2\frac{\de R_{\s\t}}{\de g^{\mu\nu}}\Phi^{\s\t}(r-s)\right\}-\frac12\,g_{\mu\nu}\cL_\Phi\nonumber\\
&\stackrel{\textrm{\tiny \Eq{giusta}}}{=}& -\int_0^{r_*}\rmd s\left\{2\Phi_{\s(\mu}(r)\Phi_{\nu)}^{\ \ \s}(r-s)-\frac{1}{D-2}\left[\Phi_{\mu\nu}(r)\Phi(r-s)+\Phi(r)\Phi_{\mu\nu}(r-s)\right]\right\}\nonumber\\
&&-\int_0^{r_*}\rmd s\left[\B\Phi_{\mu\nu}(r-s)+g_{\mu\nu}\N^\s\N^\t\Phi_{\s\t}(r-s)-2\N^\s\N_{(\mu}\Phi_{\nu)\s}(r-s)\right]\nonumber\\
&&-\frac12\,g_{\mu\nu}\cL_\Phi\,.
\ea
Including the energy-momentum tensor, we get \Eq{lasto}.

\medskip

\noindent {\bf Open Access.} This article is distributed under the terms of the Creative Commons Attribution License (\href{https://creativecommons.org/licenses/by/4.0/}{CC-BY 4.0}), which permits any use, distribution and reproduction in any medium, provided the original author(s) and source are credited.


\end{document}